\documentclass[12pt]{iopart} 
\usepackage{epsfig} 
 
\begin{document} 
\title{Hadrons in the Nuclear Medium} 
\author{M~M~Sargsian$^1$, J~Arrington$^2$, W~Bertozzi$^2$, W~Boeglin$^1$, C~E~Carlson$^4$, 
        D~B~Day$^5$, L~L~Frankfurt$^{6}$, K~Egiyan$^{7}$, R~Ent$^{8}$, 
        S~Gilad$^2$, K~Griffioen$^4$, D~W~Higinbotham$^{8}$, 
        S~Kuhn$^9$, W~Melnitchouk$^8$, G~A~Miller$^{10}$,    
        E~Piasetzky$^{6}$, S~Stepanyan$^{8,9}$,     
        M~I~Strikman$^{11}$ and L~B~Weinstein$^{9}$} 

\address{$^1$\ Florida International University, University Park, FL, USA} 
\address{$^2$\ Argonne National Laboratory, Argonne, IL, USA} 
\address{$^3$\ Massachusetts Institute of Technology, Cambridge, MA, USA} 
\address{$^4$\ College of William and Mary, Williamsburg, VA, USA} 
\address{$^5$\ University of Virginia, Charlottesville, VA, USA} 
\address{$^6$\ Tel Aviv University, Tel Aviv, Israel} 
\address{$^7$\ Yerevan Physics Institute, Yerevan, Armenia} 
\address{$^8$\ Thomas Jefferson National Accelerator Facility, 
               Newport News, VA, USA} 
\address{$^9$\ Old Dominion University, Norfolk, VA, USA}    
\address{$^{10}$\ University of Washington, Seattle, WA, USA} 
\address{$^{11}$\ Pennsylvania State University, University Park, PA, USA} 
%\ead{sargsian@fiu.edu} 

\begin{abstract} 
Quantum Chromodynamics, the microscopic theory of strong interactions,  
has not yet been  applied to the calculation of  nuclear wave functions. 
However, it certainly provokes a number of specific questions  
and suggests the existence of novel phenomena in nuclear physics which are 
not part of the the traditional framework of the meson-nucleon description of nuclei. 
Many of these phenomena  are related to high nuclear densities and the role of  
color in nucleonic interactions.  Quantum fluctuations in  the spatial separation between nucleons may lead to 
local high density configurations of cold nuclear matter in nuclei, up to 
four times larger than typical nuclear densities. 
We argue here that experiments utilizing the higher energies available upon completion of the Jefferson Laboratory energy upgrade will be able to probe  
the quark-gluon structure of such high density configurations and 
therefore elucidate the fundamental nature of nuclear matter. 
We review three key experimental programs: quasi-elastic electro-disintegration  
of light nuclei, deep inelastic scattering from nuclei at $x>1$, and  
the measurement of tagged structure functions. These interrelated programs are all  aimed at the exploration  
of the quark structure of  high density nuclear configurations. 
 
The study of the QCD dynamics of  elementary hard processes 
is another important research direction and nuclei provide a unique avenue to explore
 these dynamics. In particular, we  argue that the  
use of nuclear targets  and large values of momentum transfer 
at energies  available with  the Jefferson Laboratory upgrade
would allow us to determine whether the physics of  
the nucleon form factors is  dominated by spatially small configurations of three 
quarks. Similarly, one could determine if hard two-body processes such as   
exclusive  vector meson electroproduction are  dominated by  production of mesons  
in small-size $q\bar q $ configurations. 
\end{abstract} 
 
\submitto{\JPG} 
 
%\pacs{} 
 
\maketitle 
 
\section{Open questions in our understanding of nuclear structure} 
 
Quantum Chromodynamics (QCD), the only legitimate  candidate for a theory of strong 
interactions,  is a non-Abelian gauge theory with gauge group $SU(3)$ (in color space)  
coupled to quarks in the fundamental (triplet) representation. It contains 
the remarkable postulate of exact $SU(3)$ color symmetry. 
Quarks and gluons carry color and are the fundamental particles that interact 
via the color  force.  An important feature 
of this theory is that the $u$ and $d$ quarks, for the purpose of 
the strong interaction,  are regarded as  massless. In this limit,  
chiral  symmetry is   spontaneously broken in the ground state of QCD.
 
Nuclei should provide an excellent testing ground for this theory because 
nuclei are stable systems, made up of quarks and gluons bound together by the 
strong force.  However, the quarks and gluons are hidden and nuclei seem to 
be composed on hadrons bound together by the exchange of evanescent mesons. 
The hadrons that are the constituents of nuclei are identified with color 
singlet states and have strong interactions very different in nature than that 
of gluon exchange by colored quarks and gluons. 
 
There seems at first glance to be a contradiction between the fundamental theory,  
QCD, and the very nature of nuclei. This contradiction can be resolved by accounting 
for the fact that effective theories of the strong interaction, in which the 
degrees of freedom are hadrons, can yield results that are equivalent, 
over  some range of kinematic resolution, to those of QCD. Indeed, it is widely realized  
that the phenomenon of the  spontaneously  broken chiral symmetry in   
QCD is equivalent to the pseudovector pion-nucleon 
interaction, which accounts for the long-range, low momentum transfer, aspects of  
nuclear physics. The same chiral structure also accounts  for the relatively weak  
soft pionic fields in nuclei. Furthermore, at low momentum transfer the  
relevant degrees of freedom are  quasi-particle excitations of the system described 
by  Landau-Fermi liquid theory. 
Thus, while QCD is the fundamental theory of strong interactions, it is not 
required to explain the structure of nuclei observed in low-energy processes. 
 
The availability of high energy beams provides the opportunity to 
observe features of nuclear structure at small distance scales, which may 
reveal the presence of QCD as the ultimate source of the strong interaction. 
The central features of QCD, quarks and color, lead to two separate, but related, 
avenues of experimental investigation which we examine in this review. 
 
The first avenue is that of high density nuclear matter. What 
happens during the brief intervals when two or more nucleons overlap in space? 
Can we account for the interactions using meson exchanges, or do we 
instead  need to consider explicitly quark aspects such as  
quark exchanges between nucleons and the kneading of the nucleon's 
constituents into six- or nine-quark  bags? 
At high densities, can we detect the presence  of superfast quarks, 
i.e. quarks carrying a light cone momentum fraction greater than that 
of a nucleon at rest? 
 
The second avenue is concerned with the role of color in high momentum transfer  
processes which are exclusive (or sufficiently exclusive) so that the   
interference effects of quantum mechanics must be taken into account.  
Then, the use of high momentum kinematics offers a  remarkable 
opportunity to observe the color-singlet fluctuations of a 
hadron in  configurations of very small spatial extent, {\it viz.}, 
point-like configurations. If such  configurations exist, they  do not interact 
strongly because the effects of the emitted gluons cancel, just as the 
interaction of an electric dipole decreases with size of dipole
due to the cancellation of the electromagnetic interactions. 
Such effects are called color-coherent phenomena and include the subject 
known as color transparency. Thus, we are concerned with studying the diverse 
subjects of high-density cold nuclear matter and color coherence. 
 
While these subjects are connected by their importance in establishing QCD 
as the underlying theory of the strong interaction, they may also come 
together in the understanding of exotic phenomena in nature. 
Studies of the local high-density fluctuations are important to understanding the 
equation of state of cold, dense matter which is crucial in understanding the 
possibility of the transition of neutron stars to more dense states like  
``quark stars'', whose experimental existence was suggested recently~\cite{Drake}. 
The phenomenon of color transparency means that for a short period of time  
a nucleon can be in a spatially small configuration. In this configuration, the 
nucleon can tunnel through the potential barrier given by the repulsive core of the 
nucleon-nucleon ($NN$) interaction.  This provides a possible mechanism for a phase 
transition to a new form of matter for a sufficiently dense nuclear system, 
such as in the core of neutron star. 
 
The following is a brief outline of this review. The quark-gluon 
physics of high density fluctuations in 
nuclei is examined in Sec.~2, which begins with a brief discussion of the 
relevant history.  The discovery of the nuclear EMC effect almost twenty years ago 
brought the subjects of quarks into nuclear physics with great impact. However, 
the specific causes of the modifications observed in nuclear structure functions have 
not yet been identified with certainty. Thus, the  
questions regarding quark dynamics of nuclei  
raised by that momentous discovery have not yet been answered. The various 
ideas  invoked to explain this effect are reviewed. We then argue that an 
experimental program focused on discovering  
scaling in deep inelastic scattering at high values of  
Bjorken-$x$ ($>$1.2) will reveal the nature of high-density fluctuations of cold nuclear 
matter. Further deep inelastic scattering  
measurements of backward going nucleons in coincidence with the  
 outgoing electron offer the promise of finally determining the cause of 
the EMC effect. Section~3 is concerned with the role of color in nuclear 
physics. Electron scattering experiments at high momentum transfer in which one 
or two nucleons are knocked out, performed with sufficient precision to verify 
that no other particles are produced, could reveal the fundamental nature of color 
dynamics: that, for coherent processes, the effects of gluons emitted by a small color 
singlet object are canceled.  In that case, the dynamical nature   of the 
nucleon form factor at high momentum transfer will be revealed. 
Similarly motivated experiments in which a vector meson is produced in a coherent reaction 
with a deuteron target at large momentum transfers are also discussed.  
Section~4 presents a summary of the plan proposed to investigate the questions 
arising from the study of hadrons in the nuclear medium.

\section{Quark-gluon Properties of Superdense Fluctuations of Nuclear Matter}

\subsection {A Brief  History and Short Outlook} 
 
We begin this section with  a brief discussion of the history 
of experimental lepton-nuclear  physics. Prior to the completion of Thomas Jefferson National Accelerator Facility (Jefferson Lab), experimental studies of  nuclei using lepton probes  
could be discussed in terms of  two clearly different classes: 
 
(a) Experiments performed at electron machines with low incident electron energies, 
$E_{inc}\leq 1$~GeV, in which the typical energy and momentum transfers, $\nu$ and $\vec q$, 
were comparable to the nuclear scale 
\begin{equation} 
\nu \leq 100~\mbox{MeV}, \vert \vec q \vert \leq 2\;k_F,  
\end{equation} 
where $k_F\approx 250$~MeV/c is the characteristic Fermi momentum of nuclei. 
These reactions were inclusive $(e,e')$ and semi-inclusive $(e,e'N)$ and covered mainly 
the quasi-elastic and the low lying resonance regions (the $\Delta$ isobars), corresponding 
to relatively large values of Bjorken-$x$ ($x=Q^2/2m_p\nu$, where $Q^2=q^2-\nu^2$).

(b) Deep inelastic scattering (DIS) experiments which probed nuclei at $x<1$ and large $Q^2$  
scales, greater than about  $4$~GeV$^2$, which resolved the parton constituents of the nucleus. 
 
The first class of experiments are unable to resolve the short range structure of nuclei, and the 
second, while having good resolution, typically involved  inclusive measurements 
which averaged out the fine details and were  limited  by low luminosities and other factors. 
 
It is interesting to notice that there is a clear gap between  
the kinematic regions of these two classes of experiments.  
This corresponds exactly to the  optimal range for the study of the nucleonic  
degrees of freedom in nuclei, $1.5 \leq  Q^2 \leq 4$~GeV$^2$, for which   short-range  
correlations (SRCs) between nucleons can be  resolved, and  the  quark degrees of  
freedom are only a small  correction.  
Work at Jefferson  Lab has started to fill this gap in a 
series of  quasi-elastic $A(e,e'), A(e,e'N)$, and $A(e,e'N_1N_2)$ experiments.  
Previously, this range was just touched by  inclusive experiments at 
SLAC\cite{R1,R2,D1,D2} which also provided the first measurement of $A=2,3,4$ form 
factors at large $Q^2$. 
A number of these high-energy experiments  probe the light-cone 
projection of the nuclear wave function and in particular the 
light-cone nuclear density matrix, $\rho_A^N(\alpha,p_\perp)$, in the  
kinematics where the light-cone momentum fraction $\alpha \geq 1$  
($A\geq \alpha \geq 0$) so that short range correlations  between nucleons 
play an important role.

It is already known that the existence of  SRCs gives a natural explanation of  
the practically $A$-independent spectrum of the emission of fast backward nucleons and 
mesons, as well as the practically $Q^2$- and $x$-independent value of the 
ratio of $\sigma_A(x,Q^2)/\sigma_{D(^3He)}	(x,Q^2)$ for $Q^2\geq 1$~GeV$^2$ and 
$1.5 \leq x 
\leq 2$~\cite{fsds93,liuti93,Kim01}. Overall, these and other high-energy data  
indicate a significant probability of these correlations ($\sim$25\%  
of nucleons in heavy nuclei) which involve momenta larger than the Fermi momentum.  
These probabilities are  in qualitative agreement  with calculations of  
nuclear wave functions using realistic $NN$ potentials~\cite{Pa97}. 
Most of this probability ($\sim$80\%) is related  to two-nucleon SRCs  
with the rest  involving  $N\geq 3 $ correlations (see e.g., Ref.~\cite{FS81}). 
The current experiments at Jefferson Lab will allow a detailed study of two-nucleon  
SRCs~\cite{Arrin99,e02019}, and take a first look at three-nucleon correlations~\cite{e02019,Larry01}.  
We expect that experiments at an electron  
energy of up to $11$~GeV~\cite{wp} will allow further explorations of SRCs in the three-nucleon  
correlation region, substantially extending the region of initial nucleon momenta 
and the recoil nucleus excitation energies  that  
can be probed in quasi-elastic  $A(e,e'), A(e,e'N)$, and $A(e,e'N_1N_2)$ reactions. 
 
The 12~GeV upgrade at Jefferson Lab\footnote{The energy upgrade planned for Jefferson Lab will provide 11~GeV electrons to the experimental Halls A, B,  and C, see\cite{wp}.} will also be of a great benefit for studies of deep inelastic 
scattering  off nuclei (experiments in class (b)). It will enable us  to extend the high 
$Q^2$ inclusive $A(e,e')$  
measurement  at $x\geq 1$ to the  deep inelastic region where the process is dominated 
by the scattering from individual quarks in the nucleus with momenta exceeding the  
average momentum of a nucleon in the nucleus (superfast quarks).  
Such quarks are likely to originate from configurations  
in which two or more nucleons come close together. Thus, these measurements will  
complement the studies of SRCs by exploring their sub-nucleonic structure. 
 
The second extension of class (b) reactions concerns the measurement of  
nuclear DIS reactions in the $(e,e'N)$ semi-inclusive regime, in which the detection of  
the additional nucleon in spectator kinematics will allow us to tag  
electron scattering from a bound nucleon.  
These studies may provide a number of unexpected results similar to 
those in the inclusive studies of parton densities in nuclei, which 
yielded the observation of the EMC effect~\cite{EMC1,EMCB1,EMCB2}: 
a  depletion of the nuclear quark  parton density as compared to that 
in a free nucleon at $0.4 \leq x \leq 0.8 $, which 
demonstrated unambiguously  that nuclei cannot be described merely  
as  a collection of nucleons without any extra constituents.

Among several suggested interpretations of the EMC effect, the idea of  mesonic    
degrees of freedom  (nuclear binding models) was the only one which produced  
a natural link to the meson theory of nuclear  forces.  
This explanation naturally  led to a prediction of a large enhancement of the  
anti-quark distribution  in nuclei~\cite{dyth,dyth1,dyth2}. 
A dedicated experiment was performed  
at Fermi Lab  using the Drell-Yan process to measure the ratio 
$R_A^{\bar q}(x,Q^2)\equiv\frac{\bar   q_A(x,Q^2)}{\bar A q_N(x,Q^2)}$.   
The result of this experiment~\cite{DY} was another major surprise:     
instead of a $\sim 10-20$\% enhancement  of  
$R_A^{\bar q}(x,Q^2)$, a  few percent suppression    was observed.   
Further indications of the unexpected partonic structure of nuclei   
come from  the studies of the gluon densities in nuclei 
using exact QCD sum rules\cite{FS88,FLS}  as well as the analysis ~\cite{Pirner}  of the
data on the scaling violation of 
$R_A^{ q}(x,Q^2)$
 indicate   
that the gluon densities should be  significantly enhanced in nuclei at   
$x\sim 0.1$. 
However the inclusive nature of these measurements does not provide  
insight into the QCD mechanism for  the depletion of the DIS structure functions. 
The semi-inclusive DIS $(e,e'N)$ reactions  with the detection of backward-going nucleons will  
test many  models of the EMC effect, which was previously impossible due to  
the inclusive nature of $(e,e')$ reactions.  
Semi-inclusive DIS  reactions are  ultimately  related to the understanding of  
the QCD dynamics of multi-nucleon systems at small distances. Thus, it is crucial that  
these studies be done in parallel with the above mentioned  studies of  SRCs. 
 
\subsection{The Big Picture of Small Distance Fluctuations} 
     
The proton  electromagnetic radius  is  $\sim 0.86$ fm,  and in the ground state of 
infinite nuclear matter the average (center-to-center) distance between  
nearby nucleons is $\sim$1.7 fm. Thus, under normal conditions  nucleons are 
closely packed and nearly overlap. Despite this, quark aspects of 
nuclear structure are not evident for most of nuclear physics. A possible 
dynamical explanation for this, in terms of the strong-coupling limit of QCD, was 
presented in Ref.~\cite{scqcd}.  This can also be seen from the fact that 
typical nuclear excitations are at much lower energies than  
nucleon excitations ($\raisebox{-.4em}{$>$}\atop\sim$~500~MeV).  Besides, in  low energy
QCD, the pion, being a pseudogoldstone, interacts with the amplitude $\propto k_{\pi}$, leading to
a suppression of the near--threshold pion (multi-pion) production.
 
However, quantum fluctuations must occur in any quantum system, and well-designed 
experiments may expose the physics  occurring when nucleons occasionally  
encounter each other at smaller than average distances.  If such a fluctuation 
reduces the center-to-center separation to $1$~fm, then there is a significant  
region of overlap.  Assuming a uniform nucleon charge distribution, the density  
in the region of overlap is twice that of the nucleon, or about four to five  
times that of normal nuclear  
matter, $\rho_0=0.17$ nucleon/fm$^3$.  At these densities, the physics of confinement,  
which typifies the strong-coupling limit of QCD, may no longer be applicable and the chiral symmetry 
may be (partially) restored, see e.g. \cite{Brown}. 
One can also think of these high-density fluctuations as nuclear states with excitation 
energies large enough to modify the structure of underlying nucleonic constituents. 
Hence, dense nuclear matter 
may look very different   from a system of closely packed nucleons.    
{}From this viewpoint, it is encouraging that experimental data on the    
EMC effect indicate  that deviations from the expectations of the nucleonic  
model of  nuclei grow  approximately linearly  with the nuclear density,  
suggesting that the properties of the quark-gluon droplets could indeed  
deviate very strongly from those of a collection of  nucleons.

It is important to recall that the properties of dense  nuclear matter are  
closely related to   outstanding issues of QCD such as the existence of  
chiral symmetry restoration and  deconfinement, as well as determining the  
nature of the onset of quark-gluon degrees of freedom  and the structure   
of the phase transition from  hadronic to quark-gluon states  of  matter.   
In QCD, transitions to new phases of matter are  possible in different regimes  
of density and temperature. In particular, it has been  suggested~\cite{Wilc} 
that  nuclear matter could  exist in a   color superconductivity phase caused by the    
condensation of diquarks. Recent estimates suggest that the  average nuclear density  
could lie in between that of the dilute nucleon phase and the superconducting  
phase~\cite{Carter,Rapp}. It is therefore natural to ask whether one can observe precursors  
of such a phase transition by studying the quark-gluon properties of superdense  
droplets of nuclear matter, i.e., configurations when two or more  
nucleons come close together.  It should also be remembered that the EMC effect has been interpreted  
as a delocalization of quarks in nuclei, which is  qualitatively consistent with a 
proximity to the phase transition.  More recently, measurements of in-medium 
proton form factors also hint at such a modification of nucleon structure~\cite{Dieterich}.

\begin{figure}[ht]   
\begin{center}   
\epsfig{width=4.8in,file=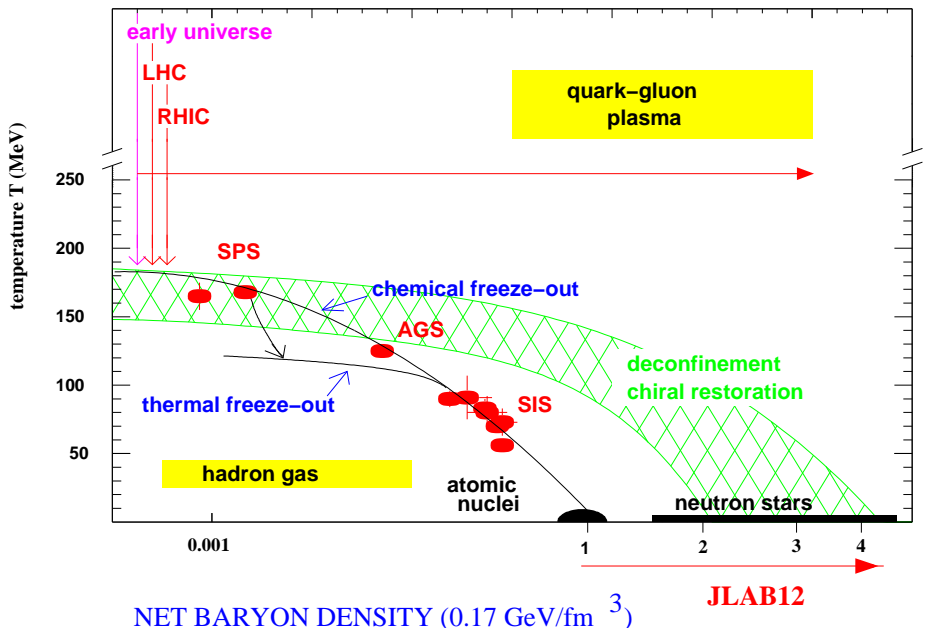}   
\end{center}   
\caption{Phase diagram for baryonic matter.}   
\label{Fig.phase}   
\end{figure}

So far, the major thrust of studies looking for the phase transitions in hadronic matter  
has been focused in the  high temperature region (Fig.~\ref{Fig.phase}), which may be realized in  
high-energy heavy-ion collisions.  The low temperature region of high densities,  
critical for building the complete picture of phase transitions  
and for determining if the transition from neutron to quark stars is possible, 
is practically  unexplored. 
We wish to argue that this  unexplored low temperature  region,  crucial for the   
understanding of the equation of state of  neutron stars, is amenable to studies  
using high energy lepton probes. Jefferson Lab, upgraded to higher energies~\cite{wp},  
would be able to explore this region and  provide studies of  nuclear fluctuations as dense as    
4--5 $\rho_0$. 
   
\subsection{Outline of the experimental program} 
 
The intellectual problems associated with establishing the existence 
of high density fluctuations of nuclear matter are formidable. We cannot  
conceive of a single experiment which would be able to answer  all of the  
interesting questions about the role of non-nucleonic degrees of freedom, 
but a series of correlated measurements may succeed.  
As mentioned above, we anticipate that studies to be done at Jefferson Lab at 
$6$~GeV will provide additional information about nucleonic degrees of  
freedom in the $2N$ sector of SRCs. This will certainly help us in refining the 
envisioned experimental program. 
 
Our view is that three key experimental programs,  for which  the use of 
$12$~GeV  electrons is essential, are needed.

\begin{itemize}   
 
\item Quasi-elastic electro-disintegration of light nuclei at 
      $1.5\leq Q^2\leq 4$~GeV$^2$, covering a wide range of angles and  
      missing momenta\footnote{We define the missing momentum as the difference 
      between momenta of knocked-out nucleon and virtual photon.} 
      (in excess of $1$~GeV/c), with the goal of 
      probing the structure of SRCs. 
 
\item  Deep inelastic scattering at $x>1$, with the  goal of  observing  
       superfast quarks in nuclei.   
 
\item Tagged structure functions (measurement of a nucleon from the  target 
      fragmentation region in coincidence with the outgoing electron) with the goal 
      of directly observing the presence of  non-nucleonic degrees of freedom    
      in droplets of superdense matter.   
 
\end{itemize}

The first goal of studying  SRCs using  quasi-elastic electro-disintegration of  
light nuclei is based on the capabilities of Jefferson Lab to perform fully  
exclusive measurements. The exclusive reactions using a deuteron target in a wide 
range  
of recoil nucleon momenta and angles will allow us to obtain insight into  
many fundamental questions, e.g. the probability of pre-existing non-nucleon  
components in the deuteron wave function, the transition of meson currents  
to the quark-interchange mechanism in the $NN$ interaction, and the dynamical  
structure of the nuclear core.  
For heavier ($A=3,4$) nuclei, the capability to measure a recoil  
nucleon with light  cone momentum fraction $\alpha \geq 2 $ will  
provide direct access to the three-nucleon correlation  
structure of the nuclear wave function providing the unprecedented opportunity  
to study the dynamics of the three-nucleon force.

The second goal of learning about superfast quarks will  be accomplished by  
studying  inclusive electron nucleus scattering in the scaling region at $x\geq   
1$, where this process measures, {\it in a model independent way}, the probability  
of finding a  quark that carries a larger light-cone momentum than   
one would expect from a quark in a low-momentum nucleon ($k \leq k_F$).   
Obviously, very few such superfast quarks can originate from the effects of the  
nuclear mean field, but the uncertainty principle  indicates that  such  quarks  
in nuclei could arise from fluctuations of  superdense configurations consisting  
either of a few overlapping  nucleons  with large  momenta, or of  more complicated  
multi-quark configurations.  The study of the structure of these superdense fluctuations    
would  be the major goal  of deep inelastic scattering measurements at high $x$. 
For instance, 
high precision  measurements of the $A$-dependence of the cross section would permit 
the investigation of the density dependence  of the superfast quark  
probability in the nuclear medium. 
The functional dependence of this probability on Bjorken-$x$  would allow  
us to disentangle the contributions of two- or multi-nucleon  
droplets from more exotic components (e.g., 6-, or 9-quark bags).

The third goal of determining the modification of the quark-gluon  
structure of nucleons in a dense nuclear  environment could be realized by 
measuring tagged structure functions in semi-inclusive DIS reactions.  
For example, the observation of a backward-moving 
nucleon or nucleons  will be used as a tag to select hard scattering    
from a short-range correlation of nucleons while the DIS characteristics of the   
event will be used to study modifications of the parton density.   
These studies primarily require  the isolation of a backward-moving nucleon  
for which a tagging procedure, rather than reconstruction of a missing nucleon, 
will provide the cleanest signature.

Successful completion of these programs will allow significant progress in   
understanding such unresolved topics of strong interaction physics as,     
{\em interaction of quarks at high densities and low temperatures}, and 
{\em the quark-gluon structure of nuclei}.

\subsection{Nucleonic Structure of Short-Range Correlations} 
 
The first step of the program will be to study the structure of SRCs 
in terms of nucleonic degrees of freedom, and map out the strength of 
two-nucleon and multi-nucleon correlations in nuclei.  This can be accomplished 
by combining moderate energy ($4-6$~GeV) inclusive scattering experiments from 
light and heavy nuclei with moderate and higher energy coincidence reactions on 
few-body nuclei, as described in the previous section.  Here we present an 
experimental overview of these reactions.

\subsubsection{Qualitative Features of High-$Q^2$ Electro-Nuclear Reactions} 
 
The observation of SRCs in nuclei has long been considered one of the most significant aims 
of nuclear physics.  These correlations, though elusive, are not small: 
calculations of nuclear wave functions with  realistic $NN$ potentials consistently  
indicate that in heavy  nuclei about $25\%$ of the nucleons have momenta above the  
Fermi surface~\cite{Pa97}. This corresponds to about $50\% $ of the kinetic energy  
coming from SRCs. The problem has been a lack of  experimental data at high-momentum transfer  kinematics  
which could decisively discriminate between the effects of the SRCs in the initial state  and the  
long-range multi-step effects such as final state interactions (FSIs) and meson exchange currents (MECs). 
Before we can proceed to study SRCs in nuclei, we must consider these processes.  In the following 
sections, we will show that some of these effects are suppressed at high $Q^2$, while the  remaining effects  do not mask the dynamics of the SRCs. 
 
\medskip 
\medskip 
 
\noindent {\it  Final State Interactions:}\\ 
Although final state interactions in nucleon knockout reactions do not disappear at 
large $Q^2,$ two important simplifications occur which make the   
extraction of information about the short-range nuclear structure possible: 
\begin{itemize} 
\item  In  high energy kinematics a new (approximate)  conservation law exists 
- the light-cone  momentum fractions of slow nucleons do not change if 
they  scatter elastically with the ejected nucleon which maintains its  
high momentum during the rescattering~\cite{FSS97,SM}. 
 
To demonstrate this feature let us consider the propagation of a fast nucleon with four-momentum  
$k_1= (\epsilon_1,0,k_{1z})$ through the nuclear medium. We chose the $z$-axis  
in the direction of ${\bf k_1}$ such that  
${k_{1-}\over m}\equiv {\epsilon_1-k_{1z}\over m} \approx {m\over 2k_{1z}} \ll 1$.  
After this nucleon makes a  small angle rescattering in its interaction 
with a bound nucleon of four-momentum $p_1=(E_1,p_{1\perp},p_{1z})$, it 
maintains  its high momentum and leading  $z$-direction  
with the four-momentum $k_2=(\epsilon_2,k_{2\perp},k_{2z})$, where  
${k_{2\perp}^2\over m^2_N}\ll 1$.  
The bound nucleon four-momentum  becomes $p_2=(E_2,p_{2\perp},p_{2z})$.  
The energy momentum conservation for this scattering allows us to write for  
the ``$-$'' component ($p_-=E-p_z$): 
\begin{equation} 
k_{1-} + p_{1-} = k_{2-} + p_{2-}. 
\label{claw} 
\end{equation} 
{}From Eq.(\ref{claw}) for the change of the ``$-$'' component of the bound  
nucleon  momentum one obtains 
\begin{equation} 
{\Delta p_{-}\over m} \equiv  {p_{2-}-p_{1-}\over m}\equiv \alpha_2- \alpha_1   
= {k_{1-}-k_{2-}\over m}\ll 1, 
\end{equation} 
where we define $\alpha_i = {p_{i-}\over m}$ ($i=1,2$) and use the fact that  
${k_{2\perp}^2\over m^2_N},{k_{1\perp}^2\over m^2_N} \ll 1$. 
Therefore $\alpha_1 \approx \alpha_2$. The latter indicates that, with an increase of  
energy, a new conservation law  emerges in which   
the light-cone  momentum fractions of slow nucleons, $\alpha$,   
are conserved. The unique simplification of the high energy rescattering is 
that although both the energy and momentum of the bound nucleon are distorted due to  
rescattering, the combination $E-p_z$ is not.  
 
\begin{figure}[t]   
\vspace{-0.4cm}   
\begin{center}   
\epsfig{angle=0,width=4.2in,file=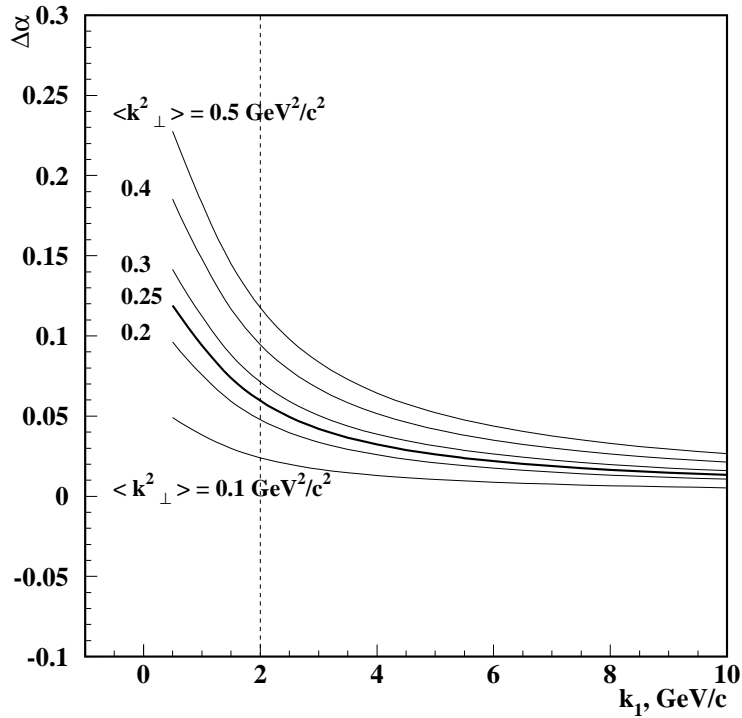}   
\end{center}   
\caption{The accuracy of the conservation of $\alpha$ as a function of the  
propagating nucleon momentum, $k_1$ at different values of average transferred  
(during the rescattering) momenta, $<k^2_t>$. Note that the eikonal 
approximation is theoretically justified for $k_1\geq 2$~GeV/c.}  
\label{Fig.alpha_conservation}   
\end{figure} 
 
Figure~\ref{Fig.alpha_conservation} demonstrates the accuracy of this conservation law for 
a propagating nucleon over a range of four-momenta relevant to our discussion. 
It is important to note that the average transferred  
momentum in the $NN$ rescattering amplitude for $p_{N}\sim 3-10$~GeV/c  
is $\langle k^2_t\rangle \approx 0.25$~(GeV/c)$^2$. Thus, starting from $3$~GeV/c momenta of  
the propagating nucleon, the conservation of  
$\alpha$ ($\sim {\cal O}(1)$)  is accurate to better than $5\%$ and improves 
with  increasing momentum. Note that the conservation of $\alpha$  to this level is  sufficient for studying  
SRCs for which the  $\alpha$ distribution of the nucleons has a rather  
slow, ($\propto \exp (-\lambda \alpha),\lambda \sim 7$) variation. 
Indeed, this variation is expected to be much flatter than the  
corresponding distribution generated  by a mean-field interaction.

\item The small angle rescatterings of high-energy ($2\leq p_N \le 10$~GeV/c) nucleons can be  
described by the  
generalized  eikonal approximation~(GEA)  which takes into account the  difference  
between the space-time picture of the proton-nucleus scattering (a proton  
coming from $-\infty$) and $A(e,e'p)$ process (a proton is produced inside the nucleus),  
and also accounts for the  non-zero Fermi momenta of rescattered nucleons~\cite{FSS97,SM}. 
Additionally, the description of small angle rescattering is simplified  
due to approximate energy independence of the $pp$ and $pn$ total cross sections in the high $Q^2$  
limit (starting at $Q^2\geq 2$~GeV$^2$, which corresponds to $p_N\geq 2$~GeV/c). 
\end{itemize} 
 
The above two features of small angle rescattering in the high $Q^2$ domain  make it possible 
to  evaluate FSIs  reliably, identifying kinematic requirements which will allow us  
to separate SRC effects from long range FSI contributions. 
 
Note that the minimal value of $Q^2$ for which one expects the eikonal approximation to be valid  
can be estimated from the application of Glauber theory to $pA$ reactions.  
Extensive studies  have demonstrated long ago that the  
Glauber theory of $pA$ processes can describe the data within a few percent   
starting at energies as low as $E^{inc}_p\geq 0.8-1$~GeV, which corresponds to  
$Q^2\geq 1.5$~GeV$^2$  in $(e,e'N)$ and ($e,e'N_1N_2$) reactions. 
 
\medskip 
\medskip 
 
\noindent{\it Contribution of Meson Exchange Currents:}\\ 
The major problem we face in the estimation of MECs in $A(e,e'N)$ processes 
is that with an  increase of energies  the virtuality of the exchanged mesons grows  
proportional to $Q^2$ ($\gg m_{meson}^2$). Even though the idea of deeply virtual  
exchanged mesons is highly complicated or may even be meaningless  
(see discussion in Ref.~\cite{Feynman}), one can still estimate its $Q^2$-dependence  
as compared to the SRC contribution.  
 
In a kinematic setting typical for studies of SRCs, in which the knocked-out nucleon  
carries almost the entire momentum of the virtual photon  (while the missing four-momentum  
of the recoil system does not change with $Q^2$), the $Q^2$-dependence of the MEC amplitude  
can be estimated as follows: 
\begin{eqnarray} 
A_{MEC}^\mu & \sim &  \int d^3p\cdot \Psi(p){J_m^\mu(Q^2)\over  
(Q^2+m_{\mathrm{meson}}^2)}\Gamma_{MNN}(Q^2) 
\nonumber \\ 
& \propto & \int d^3p\cdot \Psi(p)\left({1\over (Q^2+m_{\mathrm{meson}}^2)^2  
(1+Q^2/\Lambda^2)^2}\right), 
\label{MEC} 
\end{eqnarray} 
where $J_m^\mu(Q^2)$ is the meson electromagnetic current proportional  
to the elastic form factor of the  meson $\sim {1\over Q^2 +m_{\mathrm{meson}}^2}$, and  
$m^2_{\mathrm{meson}}\approx 0.71$~GeV$^2$. For the meson-nucleon vertices, $\Gamma_{MNN}(Q^2)$, 
we assume a  dipole  dependence with   $\Lambda\sim 0.8-1$~GeV$^2$.  
 
Since the leading $Q^2$-dependence in the SRC contribution comes from the  
nucleon elastic form factors, Eqn.~(\ref{MEC}) will result in  
an additional $\sim(1+Q^2/\Lambda^2)^{-2}$ suppression of the MEC amplitude as  
compared to the SRC contribution. Note that this gives an upper limit to the 
MEC contribution, since for large $Q^2$ the quark counting rule predicts  
stronger $Q^2$ suppression for $\Gamma_{MNN}(Q^2)\sim {1\over Q^6}$\cite{FKAS}.  
Thus, one expects that MEC contributions will be strongly  
suppressed as soon as $Q^2$ is greater than $m^2_{\mathrm{meson}}$ and $ \Lambda^2$, both $ \sim 1$~GeV$^2$. 
This conclusion is relevant only for small angle nucleon knock-out kinematics  
(optimal for studies of SRCs). In the large angle kinematics quark-exchange mechanism 
may become important similar to the large angle $\gamma d\to pn$ reactions\cite{gdpn}. 
 
\medskip 
\medskip 
 
\noindent 
{\it Isobar Current (IC) Contribution:}\\ 
For the case of IC contributions, the virtual photon produces the  
$\Delta$ isobar in the intermediate state which subsequently  
rescatters off the spectator nucleon through the  $\Delta N \to NN$  
channel. There are several factors which contribute to the suppression of  
IC contributions at high $Q^2$ as compared to the SRC contributions.  
The main factors which should be emphasized are the energy dependence of the  
$A_{\Delta N\to NN}$ amplitude and the $Q^2$-dependence of the  
electromagnetic $\gamma^*N\Delta$ transition form factors, as compared with the   
elastic $NN\to NN$ amplitude and the $\gamma^*N$ form factor respectively. 
 
The  $\Delta N\to NN$ amplitude is known to be dominated by the pion Reggeon exchange  
with the $\rho$-Reggeon which dominates at very high energies being a small correction 
up to the energies $\sqrt{s} \sim 30$~GeV~\cite{ISR}.  
Based on the rule that the energy dependence of the Feynman amplitude of the scattering process  
is defined by the spin, $J$  of the exchanged particle as: $A\sim s^J$ one observes  
that the  $\Delta N\to NN$ transition  amplitude is  suppressed  
at least by a factor $\propto 1/Q^2$ (at $Q^2 \geq 2$ GeV$^2$) 
as compared to the elastic $NN\to NN$ amplitude leading to a  
similar suppression for IC contribution.  In addition there is the 
experimental indication that the  
electromagnetic $\gamma^*N\Delta$ transition  is 
decreasing  faster with $Q^2$ as compared  
to the elastic $\gamma^*NN$ transition amplitude~\cite{Stoler}. 
 
\medskip 
\medskip

It follows from the above discussions that the smallest value of $Q^2$ required for effective  
studies of SRCs in the discussed class of experiments is  
$Q^2\geq 1.5$~GeV$^2$. The upper limit for $Q^2$ for studies of SRCs comes 
from the onset of color coherence phenomena at $Q^2> 4$~GeV$^2$,  
when FSIs will not maintain their energy independent characteristic for small angle  
hadronic rescattering (a discussion on color coherence is given in Sec.3).  
 
Hence the optimal range for probing SRCs is  
\begin{equation} 
1.5 \leq Q^2\leq 4~ \mbox{GeV}^2. 
\label{range} 
\end{equation} 
This range is large enough to check the validity of the $Q^2$-independence of the  
extracted parameters of the SRCs. Jefferson Lab  experiments at $6$~GeV and very large missing momenta  
can reach the lower limit   
of this range and they would have only a very limited access to the upper limit  of  
Eqn.~(\ref{range}). With the upgrade of the beam energy the whole range of Eqn.~(\ref{range}) will  
be be easily accessible for Jefferson Lab. This also permits a wider coverage  of missing momenta  
and excitation energies of the recoil nuclear system.

\subsubsection{Specific Reactions for Studies of Short Range Correlations} 
 
The study of the $(e,e')$, $(e,e'N)$ and  $(e,e'NN)$ reactions in the 
($1.5 \leq Q^2\leq 4$~GeV$^2$)   
range will allow a direct measurements of the nucleon momentum distributions  and  
spectral functions out to large  momenta, between $400$ and $700$~MeV/c. 
Here one expects that the nucleon degrees of freedom are dominant and one can explore  
short-range  correlations. 
Although two nucleon correlations are expected to be the dominant 
part of the SRCs, triple and higher order SRCs (where more than two nucleons come close together) 
are significant as well - they are estimated to  constitute  
$\sim$20\% of all  SRCs~\cite{FS81}. 
At initial momenta  $> 700$~MeV/c the non-nucleonic degrees of freedom should  
play an increasingly dominant role and for the first time  one will be able to investigate  
them in detail. 
 
The prime reactions for these investigations are: 
  
{\bf $\bullet$} ~~ {\bf Inclusive {\boldmath $A(e,e')X$} reaction at {\boldmath $x>1$}.} In the kinematic  
range of  Eqn.~(\ref{range}), $A(e,e')X$ reactions at $x>1$ proceed mainly by  
quasi-elastic scattering of electrons from bound nucleons. By increasing the values of  
$x$ and $Q^2$ for these reactions, one can achieve a better discrimination between  
SRCs and long-range multi-step processes. This  will allow  the measurement of several  
average characteristics of SRCs: the probabilities of two-nucleon correlations  
in nuclei~\cite{Kim01}  and  $k_\perp$-averaged longitudinal/light-cone momentum  
distributions in SRCs~\cite{John01}. Extending measurements into the $x>2$ region at sufficiently  
large values of $Q^2$ will allow us to probe the three-nucleon correlations. The  
signature of the dominance of 3N correlations in these reactions will be the  
onset of scaling of the cross section ratio of scattering from  
nuclei with $A>3$  to that  of $A=3$ nucleus in the $x>2$ region.  Alternatively, 
one can look for multi-nucleon correlations by extending the $Q^2$ range of 
measurements for $x<2$, and comparing heavy nuclei to deuterium.  QCD evolution 
of the structure function leads to a shift of strength from high-$x$ 
to lower-$x$ values as $Q^2$ increases.  If only two-nucleon correlations are 
present, then heavy nuclei will have no appreciable strength above $x=2$ and the 
evolution should be essentially identical to the evolution of deuterium, and the ratio 
will remain constant.  If, however, there is significant strength above $x=2$ in 
heavy nuclei, coming from multi-nucleon correlations, then this strength will 
shift into the $x<2$ region, and the ratio $\sigma_A/\sigma_D$ will increase 
with $Q^2$.

{\bf $\bullet$} ~~ {\boldmath $d(e,e'pn)$}.  While inclusive reactions have a large kinematic  
reach, they cannot probe the details of the structure of the SRCs.   
The starting point for these studies is the simplest exclusive  
reaction: $e+d \to e+p+n$. It will provide a test of the current understanding of the  
dynamics of  electro-disintegration processes especially since the wave function of the deuteron  
is reasonably well known for a wide range of momenta ($\leq 400$~MeV/c). 
Progress in building tensor polarized deuteron targets makes it feasible to study  
the polarization degrees of freedom of the disintegration reaction  at sufficiently large $Q^2$.  
In this case a direct separation of $S$ and $D$ waves is possible. Hence this process will provide  
an ultimate test of our understanding of the short-distance $NN$ interactions. In particular,  
it will allow us clearly discriminate between  
predictions based on approximations to  
the Bethe-Salpeter equation and light-cone approaches to the description  
of the deuteron as a two-nucleon relativistic system~\cite{FS78,FS83}. 
Such studies, using  the upgraded energies at Jefferson Lab will be a natural continuation of the  
present experimental program of electro-disintegration of the deuteron which is currently  
focused on studies in the momentum range of $\leq 400$~MeV/c~\cite{UJ,KG,EGS,WJKUV}.

{\bf $\bullet$} ~~ {\bf {\boldmath $A(e,e'N)$} and {\boldmath $A(e,e'NN)$} with  
{\boldmath $A\geq  3$}}. 
These  reactions, with  one nucleon produced  along $\vec q$ carrying almost the entire momentum  
of the virtual photon, allow  measurements of the light-cone density matrix of the nucleus  
$\rho_A^N(\alpha,p_\perp)$ for large values of excitation energy, $E_m$, of the residual system. 
Within the SRC picture,  it is expected that $E_m$ increases with increasing   
initial  momentum of the ejected nucleon. 
In the non-relativistic approximation, the average  
excitation energy is   $\left<E_{m}\right>\approx p_m^2/2m_N$, 
where $\vec p_m$ is the missing momentum of the ejected nucleon. 
Measuring the $E_m$--$p_m$ correlation at high $Q^2$ will be one of the  
signatures of scattering from SRCs. 
 
Note that polarized $^3$He targets used in  $A(e,e'NN)N$ reactions  
will play a special role for probing SRCs due to the   
relative simplicity of the wave function  and the unique  
possibility to probe  the spin structure of $pp$ and $pn$ correlations. 
In particular, there exist kinematic regions where a minimum in  
the S-wave $pp$ wave function can be explored, and the  P-wave contribution  
can be isolated. These measurements will  provide stringent tests of  the  
structure of $A=3$ systems and  will  test current interpretations  
of measurements of the $^3$He form factors at large $Q^2$.

{\bf $\bullet$} ~~{\bf  {\boldmath $A(e,e'N_fN_b)$} reactions} with  one nucleon ($N_f$)  
moving forward and the other ($N_b$) moving  backward for nuclei with $A\leq 12$ can be  
used to investigate how the excitation energy  
is shared between nucleons. It is expected that the  dominant contribution will originate from  
two-nucleon correlations. In this case  $N_b$ should carry most of the excitation energy.  
A comparison of the yields of ($pp$), ($pn$) and ($nn$) processes will  provide a detailed check  
of the mechanisms of the reaction and provide a quantitative comparison between   
the wave functions of two nucleon SRCs in the isospin zero and one channels   
(the former is expected to dominate by a  factor $\geq 4$  
for a large range of momenta).  In addition, if there is significant strength in 
multi-nucleon correlations, it should be manifest  in the low excitation tail  
of the nuclear spectral  function for large momenta of the ejected nucleon. 
This is best  observed through the  $(e,e'NN)$  reaction in which two nucleons  
are emitted  in the backward direction relative to the virtual photon momentum~\cite{FS88}.  
In the high $Q^2$ regime, these reactions will allow us to study the parton 
structure of the three nucleon correlations at very high densities.

{\bf $\bullet$} ~~ {\bf  Quenching in the {\boldmath $A(e,e'N)$}  scattering for {\boldmath $A\geq 10$}.} 
The numerous  $A(e,e'p)$ experiments at low $Q^2\leq  0.3$~GeV$^2$ have observed  
the shell structure of nuclei for the momenta of residual nuclear system $\leq k_F$. 
At the same time they observed a significant  
($\sim 0.5 $) suppression of the absolute cross sections as compared to  
shell model expectations. 
Jefferson  Lab measurements  
on $^{12}$C~\cite{Dutta} and $^{16}$O~\cite{bert} targets at  intermediate 
$Q^2 \sim 0.6-0.8$~GeV$^2$ suggest a quenching of about a factor of 0.7--0.8. 
For large $Q^2$ ($\raisebox{-.4em}{$>$}\atop\sim$~2~GeV$^2$) where virtual photons  
resolve individual nucleons the analysis~\cite{Lap,Zhalov}  of the current data 
including the Jefferson Lab data~\cite{quench} 
indicate that this suppression has practically disappeared.
Note however that the current comparison of quenching at different 
$Q^2$ should be considered as a semi--quantitative since different models of
nucleon absorptions were used for treating $A(e,e'p)$ reactions at 
different $Q^2$.  
Hence it is  very important to perform similar measurements  
including a separation of the different structure functions at 
$Q^2\sim 2- 3$~GeV$^2$ to investigate the $Q^2$-dependence of the quenching. 
Such studies would be also of importance for the interpretation of the  
color transparency searches which will be discussed in Section 3.

Overall, a series of experiments at Jefferson Lab can provide a detailed knowledge of the  
nucleon component of the SRCs, mapping out both the strength and structure of two-nucleon 
(and multi-nucleon) SRCs for both light and heavy nuclei. 
With the proposed upgrade in beam energy, these studies  
will probe for the first time the quark substructure of the nucleon configurations 
at short space-time separations.

\subsubsection{Experimental Requirements}

Carrying out the scientific program described above involves 
inclusive $A(e,e')$ and double coincidence $A(e,e'p)$ measurements as well as triple coincidence  
measurements of the $(e,e'pN)$ reactions on light nuclei, {\it e.g.} carbon. 
Inclusive measurements can be performed over a wide range at $6$~GeV, 
and extended to the highest $x$-values with $11$~GeV. 
The coincidence measurements are to be performed at the highest incident energy ($11$~GeV) 
and momentum transfer ($Q^2=4-6$~GeV/c$^2$).  
These kinematical conditions are essential for covering the largest possible  
missing momentum range (up to $1$~GeV/c). For this we need  
an electron spectrometer similar to that of the current high resolution  
spectrometer (HRS) of Hall A at Jefferson Lab,  with extended  momentum acceptance  
up to about $10$~GeV/c, and a 
proton spectrometer with momentum acceptance up to about $2.5$~GeV/c. 
For the triple coincidence measurements we also need a third large  
solid angle proton spectrometer and a neutron array.  For protons,  
the BigBite spectrometer~\cite{BB}, which at its maximum current can detect  
particles in the  momentum range of  $250-900$~MeV/c with    
moderate momentum resolution of  $\Delta p/p=0.8\%$, can be used. 
Behind the BigBite spectrometer a  neutron counter array 
can be installed with a matching solid angle.  
This is basically the set-up proposed for an approved $(e,e'pN)$ 
experiment~\cite{add5:CEBAFpro} with the current Jefferson Lab accelerator (at $5$~GeV).

The limiting factor for luminosity  in these measurements is the singles 
rate in the large solid angle detectors used to detect the 
recoil particles in coincidence with the knockout proton. 
Assuming that for the  upgraded energies at $11$~GeV the singles rate  
will be similar to that of the current one at $5$~GeV beam,  one can use 
$100~\mu A$ beam current  and $1$~mm carbon target to obtain  the nuclear luminosity   
${\cal L} = 6\times 10^{36}$ cm$^{-2}$sec$^{-1}$. For the  
electro-disintegration of the deuteron we assume a target similar 
to the  one used in the current Jefferson Lab experiments. The luminosity 
in this case (for $100 \mu A$ beam and $15$~cm of $0.16$~gr/cm$^2$ deuteron 
target) is:  ${\cal L}=3.7\times 10^{37}$cm$^{-2}$sec$^{-1}$. 
 
For the differential cross section for the $d(e,e'p)n$ reaction, estimated in the  missing  
momentum range of  $450-500$~MeV/c in (almost) anti-parallel kinematics and at $Q^2=4$GeV$^2$, 
one obtains 
\begin{equation} 
{{d\sigma \over{dE_e d\Omega_e d\Omega_p}}_d= 1 {pb \over MeV sr^2}}. 
\label{m1} 
\end{equation} 
Assuming $\Delta E_e=50$~MeV and $\Delta \Omega_e=\Delta \Omega_p = 10$~msr, 
the counting rate   is 650 events/hour. 
With  a requirement of at least 500~events in 100~hours of beam time one will be able to measure  
the differential cross section as low as $0.005{pb\over MeV sr^2}$.  Thus 
one will be able to measure well beyond  $500$ MeV/c region of missing  
momentum in  the $d(e,e'p)n$ reaction. 
 
To estimate the triple coincidence  counting rate for the 
$A(e,e'pn)X$ reaction,  we assume that events with  
high missing momenta ($p_m>500$~MeV/c)  originate mainly from 
two-nucleon SRCs. Under this condition, the  
measured differential cross section, in which the solid angle of the
spectator neutron is integrated around the direction
defined by the deuteron kinematics, can be approximated in the following 
way~\cite{add5:CEBAFpro}:
\begin{equation}
{d\sigma^A \over{dE_e d\Omega_{e}\Omega_p}}={ K_0\times a_2\times Z}\times
{d\sigma^d \over{dE_e d\Omega_e \Omega_p}}
\label{eli1}
\end{equation}
where  $ {d\sigma^d \over{dE_e d\Omega_e d\Omega_p}}$ is the differential 
cross section of the $d(e,e'p)n$ reaction~\cite{FGMSS95}, $a_2$ is 
defined through the
ratio of cross sections for inclusive $(e,e')$ scattering from heavy 
nuclei to deuterium,
measured at $x>1$ where the scattering from SRCs dominates,  
($a_{2}(^{12}C)  = 5.0\pm 0.5$~\cite{fsds93}). $K_0$ is a kinematical 
factor related to  the
center of mass motion of $np$ correlations  in the nucleus and defined 
by the integration
range of the  spectator neutron  solid angle.
For $K_0$ we used a conservative estimate of $0.2$ based on the neutron 
detector
configuration designed for the experiment of Ref.~\cite{add5:CEBAFpro}.
Assuming additionally that the neutron detection efficiency is $50\%$ 
one obtains 
a triple $(e,e'pn)$ coincidence rate of approximately 250 events/hour.
With a requirement of  at least $500$~events in $100$~hours of 
beam time a differential cross section as low as
$0.02~{pb\over MeV sr^2}$ can be measured.

The rate for the $A(e,e'pp)X$ reaction is more difficult to estimate since  
the $pp$ correlations cannot be approximated by the high momentum part of  
the deuteron wave function. The ratio between the $np$ and $pp$ short range  
correlation  contributions is poorly known and is one of the anticipated outcomes of the  
proposed measurements. For estimation purposes we  assume that $2 \leq (np)/(pp) \leq 4$. This assumption is 
 based mainly on counting the isospin degrees of freedom.

\subsection{Quark Structure of Short-Range Correlations - 
Study of  Superfast Quarks }

The discovery of Bjorken scaling in the late 1960's~\cite{scaling} was one of  
the key steps in establishing QCD as the microscopic theory of strong 
interactions. 
These experiments unambiguously demonstrated that 
hadrons contain   point-like constituents---quarks   
 and gluons (see e.g.~\cite{Bjorken,Feynman}). 
In the language of quark-partons, the explanation 
of the observed approximate   scaling was remarkably simple: a  
virtual photon knocks out a  
point-like quark,   and  the structure function of the target 
nucleon measured  experimentally depends on the fraction $x$ of the 
nucleon light cone momentum ($\equiv +$)  that the quark carries (up to 
$\log Q^2$ corrections calculable in pQCD).    
For a single free  nucleon the  
Bjorken variable, $x= {Q^2 / 2 m_N \nu}\leq 1$. 
It is crucial that the QCD factorization theorem be valid for this process, in which case the 
effects of all initial and final   state 
interactions are canceled  and the deep inelastic  
scattering can be described in terms of a light-cone wave function of  
the nucleon.  
Experimentally such  scaling was observed, for a hydrogen 
target,  for  $Q^2 > 4$~GeV$^2$ and $W > 2$~GeV. 
Here  $W^2 = -Q^2 + 2m_N\nu + m_N^2$ is    the invariant mass squared of 
the hadronic system produced in the   
$\gamma^*$--$N$ interaction.

Since the nucleus is a loosely bound system it is natural to redefine   
the Bjorken variable, for a nuclear target,   
as $x_A={AQ^2 / 2m_A\nu}$ ($0 \leq x_A \leq A$), 
so that for scattering in  kinematics allowed for a free  
nucleon at rest $x_A \approx x$.  
In the case of electron scattering from quarks    
in  nuclei it is possible to have Bjorken--$x > 1$. 
This  corresponds to the situation that a knocked-out quark carries a    
larger  light-cone momentum fraction than a nucleon which is at rest   in 
the nucleus. Such a situation could occur, for example, if  
the quark belongs to a fast nucleon   in the nucleus.  In the impulse 
approximation picture, the DIS structure function of a nucleus, $F_{2A}(x,Q^2)$, which  
directly relates   to the nuclear quark distribution function, is expressed in terms of 
the nucleon structure function and the nuclear light-cone density matrix: 
\begin{equation}   
F_{2A}(x,Q^2)=\int_x^A   
\rho^N_A(\alpha_N,p_\perp)F_{2N}(\frac{x}{\alpha_N},Q^2)
\frac{d\alpha_N d^2p_\perp}{\alpha_N}, 
\label{conv}   
\end{equation}   
where $\alpha_N$ is the light-cone momentum fraction of the nucleus carried by  
the interacting  nucleon.  
Choosing $x\geq 1 + k_F/m_N \approx 1.2$ almost completely eliminates the contribution  
of scattering  by   quarks belonging to  nucleons with momenta  
smaller than the Fermi momentum.  Actually, the use of 
any realistic  nuclear wave function would yield the result that 
the contribution of the 
component  of the wave function with $k\geq k_F$ dominates at large $Q^2$ for values 
of $x$ as small as unity.   For these values of $x$, 
a quark must acquire  its momentum from multiple nucleons   
with large relative momenta which are significantly closer to each other 
than  the average inter-nucleon distance~\cite{FS81}.    Thus, such 
superfast quarks in nuclei could 
arise from some kind of  superdense configurations consisting either  of 
a few nearby nucleons with   large momenta or of  more complicated multi-quark 
configurations.  In particular,  a comparison with Eqn.~(\ref{conv}), which 
builds the nucleus from a distribution of essentially free nucleons, would 
provide a  quantitative test  whether the quarks in a bound nucleon have the  
same distribution function as in a free nucleon. 
 
\begin{figure}[htb] 
        \centerline{ 
        \hbox{\epsfig{figure=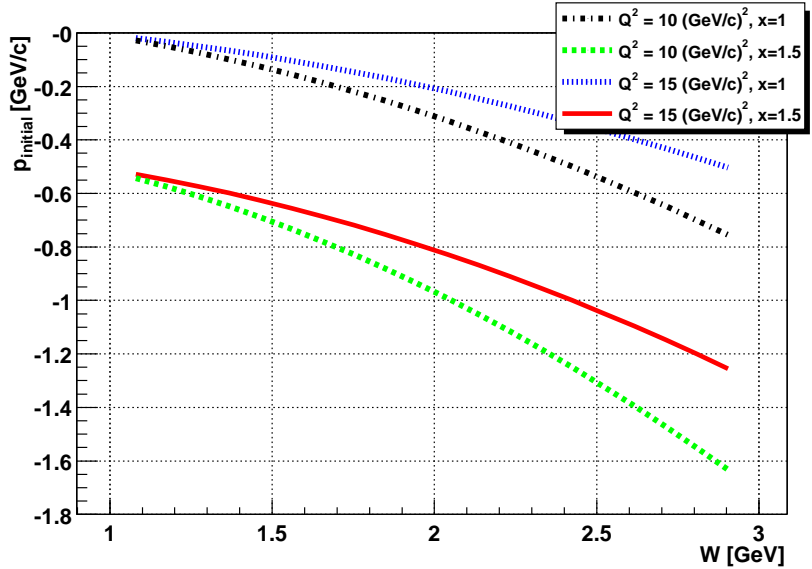,height=8cm} } } 
        \caption[]{\protect\label{fig1}Relation between $p^z_{initial}$ and 
          $W_N^2$ for $Q^2=10, 15$~(GeV/c)$^2$ and $x = 1, 1.5$} 
\label{Fig.kinw} 
\end{figure}

The kinematic  requirement  for detecting the signature of superfast quarks at  
$x>1$  is to provide a value of   $Q^2$  large enough that the tail of  
deep-inelastic scattering would  overwhelm the contribution from quasi-elastic  
electron scattering from  nucleons.  One can estimate approximately   
the magnitude of the initial momentum of the  nucleon, which is relevant  
in the DIS channel, expressing its projection in the {\boldmath $q$} direction  
through the produced invariant mass, $W_N$ associated with scattering from a  
bound nucleon: 
\begin{equation} 
{p^z_{initial} \over m} = 1-x-x\left[{W_N^2- m^2 \over Q^2}\right]. 
\end{equation} 
For a free nucleon, a given value of $x$ and $Q^2$ automatically fixes the  
value of $W_N^2$. In contrast, the internal motion of a bound nucleon 
allows a  range of values of $W_N^2$ for  given values of  
$x,Q^2$. When average $W_N\ge 2$~GeV the deep inelastic contribution 
becomes dominant and the Bjorken scaling limit is reached. 
Figure~\ref{Fig.kinw}  
demonstrates that DIS can access extremely large values of initial momenta in nuclei  
at large $Q^2$ and $x$.

\begin{figure}[htb]   
\begin{center}   
\epsfig{width=4.0in,file=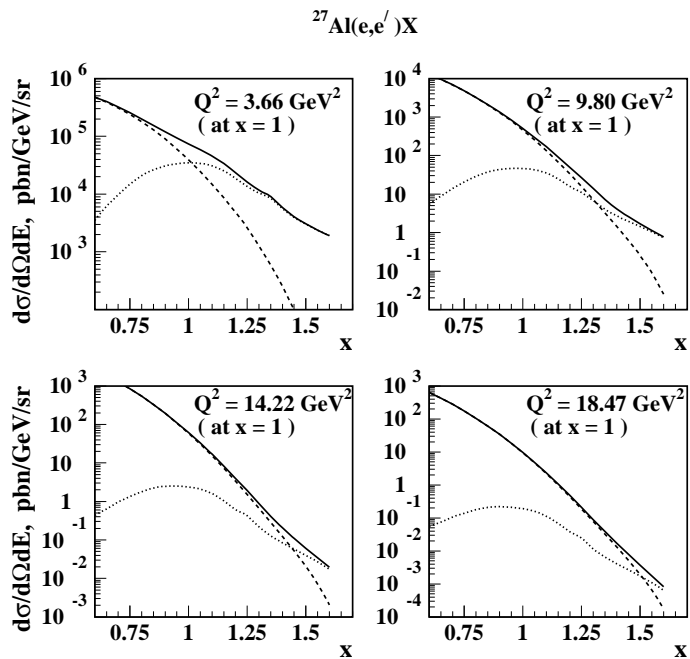}   
\end{center} 
\vspace{-0.4in} 
\caption{The differential $^{27}Al(e,e')X$ cross section as  
a function of $x$ for fixed beam energy and scattering angle. 
The dotted line is the quasi-elastic contribution, the dashed line 
is the inelastic contribution, and the solid line is the sum of both contributions. 
Values of $Q^2$ are presented for  $x=1$. }   
\label{Fig.crsal}   
\end{figure}

However, in order to probe these values of initial momenta the DIS contributions should  
dominate the quasi-elastic contribution.  
Figure~\ref{Fig.crsal} shows a calculation of the   
$A(e,e')X$ cross section  at four different values of $Q^2$. This figure 
illustrates that with  increasing $Q^2$ the inelastic contribution  
remains dominant at increasingly larger  values of $x$.

The first signal of the existence of superfast quarks will be the  
experimental observation of scaling in the region $x\geq 1$.   
Previous experimental attempts to observe such superfast quarks were  
inconclusive: the BCDMS collaboration~\cite{BCDMS} has observed a  
very  small $x\geq 1$ tail ($F_{2A}\propto \exp(-16x)$),  
while the CCFR collaboration~\cite{CCFR} observed  a tail  consistent  
with  presence of very significant SRCs ($\sim \exp (-8x)$).    
A possible explanation for the  inconsistencies is that the resolution in  
$x$ at  $x \geq 1$ of   the high energy muon and neutrino experiments   
is relatively poor,  causing great  difficulties in measuring    
$F_{2A}$ which is expected to vary rapidly with $x$.   
Therefore the energy resolution, intensity and energy of Jefferson Lab 
at $11$~GeV may  allow it to become the first laboratory to observe the onset of    
scaling and thereby confirm the existence of superfast quarks.

To estimate the onset of the Bjorken scaling we should extract the  
structure functions $F_{2A}$ and $F_{1A}$ from the cross section of the 
inclusive $A(e,e')X$ reaction. The cross section can be represented    
as follows:   
\begin{equation}   
{d\sigma_A\over d\Omega_e dE_e'} =    
{\sigma_{\rm Mott}\over \nu} \left[F_{2A}(x,Q^2) +    
{2\nu\over m_N} \tan^2(\theta/2)F_{1A}(x,Q^2)\right],   
\label{crs2}   
\end{equation}    
where $F_{2A}(x,Q^2)$ and $F_{1A}(x,Q^2)$ are two invariant   
structure functions of nuclei. In the case of scaling, both    
structure functions become independent of  $Q^2$ (up to $\log Q^2$ terms).

\begin{figure}[ht]   
\vspace{-0.4in} 
\begin{center}   
\epsfig{width=4.0in,file=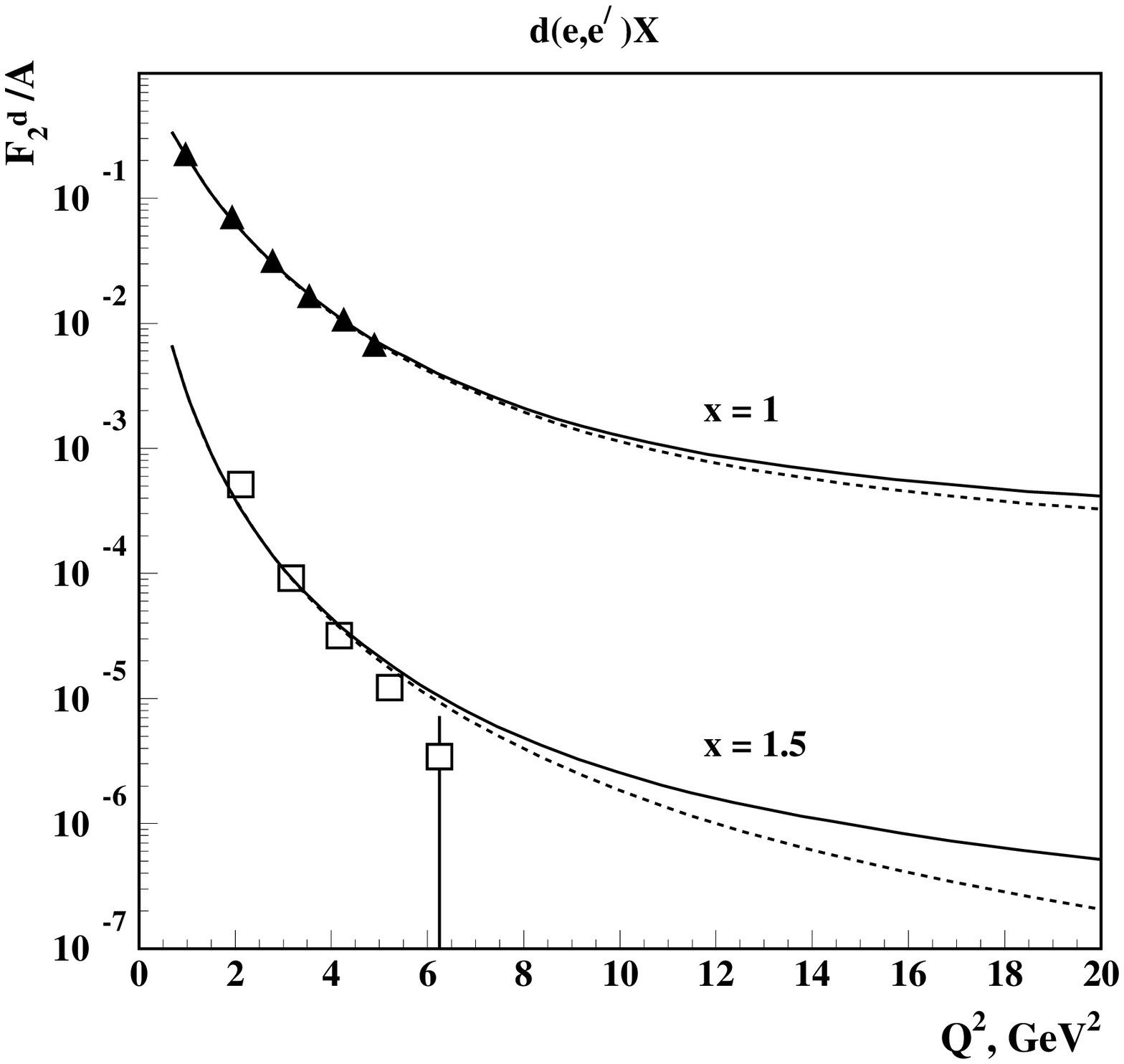}   
\end{center} 
\vspace{-0.6in} 
\caption{\protect\label{Fig.f2d}$F_{2d}/A$ as a function of $Q^2$ for $x=1$ and $x=1.5$.  
Solid lines correspond to the light-cone calculation of Ref.~\cite{FS88} with  
no modification of the nucleon structure function.   
Dotted lines account for the binding modification of the nucleon DIS structure  
functions within the color screening model discussed in Sec.~\ref{sec:emcmodels}. 
No binding modification  of elastic structure functions are considered. 
The data are from Ref.~\cite{Arrington}.} 
\end{figure}

\begin{figure}[ht] 
\vspace{-0.6cm} 
\begin{center} 
\epsfig{width=4.2in,file=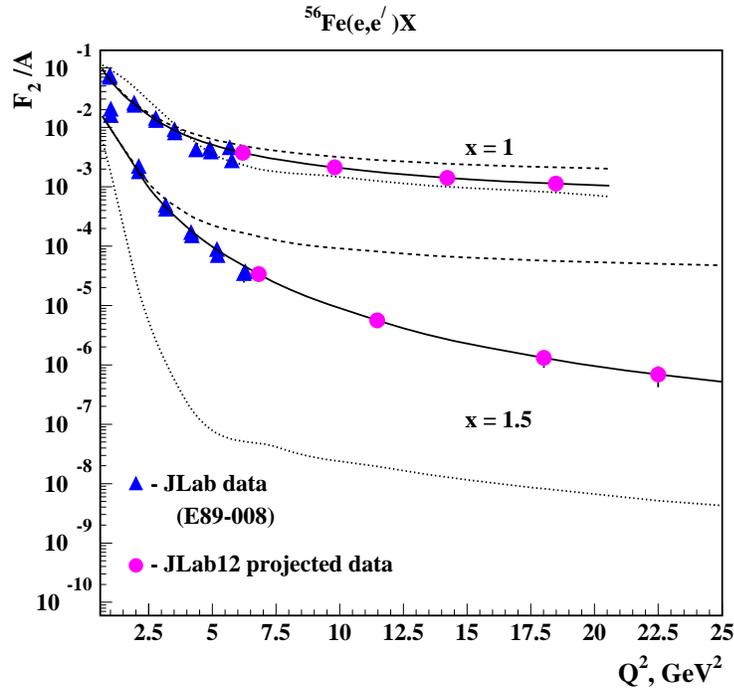} 
\vspace{-0.8cm} 
\end{center} 
\caption{\protect\label{Fig.fescale}Prediction of the onset of  scaling for $^{56}$Fe$(e,e')X$  
scattering. The data are from Ref.~\cite{Arrin99} and the curves are 
the mean-field (dotted), two-nucleon SRCs (solid) and multi-nucleon 
SRCs (dashed) models, as described in the text.} 
\end{figure}

The experimental observable for scaling is the structure function, $F_{2A}$: 
\begin{equation}   
F_{2A}(x,Q^2) = {{d\sigma_A\over d\Omega_e dE_e'}   
\left( {\nu\over \sigma_{\rm Mott}}\right)\left[1 + {1-\epsilon\over \epsilon}    
{1\over 1 + R(x,Q^2)}\right]^{-1}},   
\label{fnkar}   
\end{equation}   
 
where  $\epsilon = [1 + 2(1+{\nu^2 / Q^2})\tan^2({\theta / 2})]^{-1}$  
and $R \equiv {\sigma_S / \sigma_T}=  (F_{2A} / F_{1A})(m_N / \nu)(1+{\nu^2\over Q^2})-1$. 
Figures~\ref{Fig.f2d} and ~\ref{Fig.fescale} display  calculations~\cite{FDSS,FSS90}  
for  deuteron and iron targets. Figure~\ref{Fig.f2d} demonstrates that at high $Q^2$  
calculations become sensitive to the binding modification of the DIS structure function of  
the nucleons. To appreciate the size of the possible modification we used one of the  
models (color screening model of Ref.~\cite{FS85}) which describes reasonably well the  
nuclear EMC effect at $x<1$ (see Sec.~\ref{sec:emcmodels} for details). 
The calculations for $^{56}$Fe show that the onset of scaling  
($Q^2$-independence) at $x=1$ is expected at values of $Q^2$ 
as low as $5-6$~GeV$^2$. At $x=1.5$ the onset of scaling  
depends strongly  on the underlying model of SRCs and may occur already at  
$Q^2\sim 10$~GeV$^2$.  
 
Figure~\ref{Fig.fescale} shows results obtained using  three different models  
describing the ($A>3$) nuclear state  containing the superfast quark.    
In the first model,  the momentum of the target nucleon in the   
nucleus is assumed to be generated by a  
mean field nuclear interaction only (dotted line).  In the second,   
the high momentum component of the  
nuclear wave function is calculated using a  
two-nucleon short range correlation model 
(solid lines).  Within this approximation the  variations  of the structure 
functions with $x$ will 
be the same for  deuteron and $A>2$ targets at large values of $Q^2$. 
 In the third model, the multi-nucleon correlation model (dashed 
lines)  of Ref.~\cite{FS80} is used.  This model agrees reasonably well with recent  
measurements of the   nuclear structure functions by  the  
CCFR collaboration~\cite{CCFR} but yields a significantly larger  
quark distribution than the one reported by the BCDMS collaboration~\cite{BCDMS}.   
     
\begin{figure}[ht] 
\vspace{-0.6cm} 
\begin{center} 
\epsfig{width=4.0in,height=3.5in,file=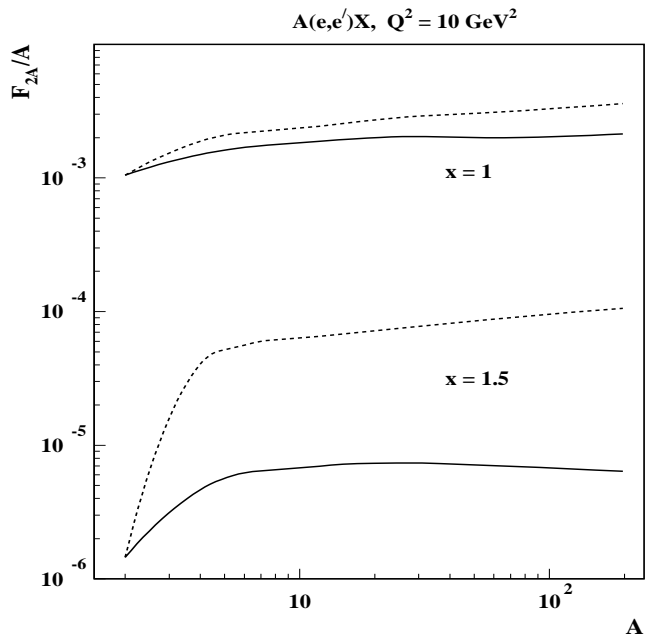} 
\end{center} 
\vspace{-0.6cm} 
\caption{A-dependence of the structure function.  The solid curve includes  
only two-nucleon SRCs, while the dashed curve includes multi-nucleon SRC contributions.} 
\label{Fig.adep} 
\end{figure}

Figure~\ref{Fig.adep} represents the $A$-dependence of $F_{2A}$, which  
emphasizes that the use of large nuclei and large values of $x$ would   
allow the significant study of the models  of short-range correlations.

\subsubsection {Experimental Requirements}

The physics program described above involves extending  
current inclusive scattering measurements at Jefferson Lab to the  
highest possible values of $Q^2$ at $x\geq 1$. The extension 
to higher $Q^2$ values requires the detection of high momentum electrons over a  
wide angular range $\theta\leq 60^\circ$. To reach the largest possible $x$ values, we need 
to detect extremely high energy electrons at angles up to 30$^\circ$. 
In both cases the measurements will require  high luminosity, excellent pion  
rejection and a moderate ($\sim10^{-3}$) momentum resolution. 
We use the measurement of the structure function $F_2(x,Q^2)$ for nuclei, as 
well as the ratio between Aluminum and Deuterium, 
as an example of a possible experiment that can be carried out with the Jefferson Lab  
upgrade.  The extension to the highest possible $Q^2$ values, 
necessary to reach scaling and probe the quark distributions, can be 
performed using the equipment proposed for either Halls A or C at Jefferson 
Lab.  The extension to the highest possible $x$ values, where sensitivity 
to multi-nucleon correlations is greatest, requires the detection of 
electrons with energies approaching the beam energy, and will require 
a very high momentum spectrometer, such as the proposed SHMS in Hall C. 
 
In order to estimate the feasibility of these measurements at large 
values of $x$ we have estimated count rates for Deuterium and 
Aluminum targets.  
The incident beam energy is assumed to be $11$~GeV and a beam  
current of $60 \mu A$ is used.  
We used a $10$~cm long Deuterium target which corresponds to 1.5\% of a 
radiation length. This implies a luminosity of $1.9\cdot 
10^{38}$s$^{-1}$cm$^{-2}$. These luminosities are currently used in many 
experiments at Jefferson Lab and do not pose any technical problems. 
For the Aluminum target we assumed a thickness of $0.5$~cm which 
corresponds to a 6\% radiator. This target will also have to be 
cooled but this should not pose any special problems. For the 
Aluminum target we then obtain a luminosity of 
$1.2 \cdot 10^{37}$s$^{-1}$cm$^{-2}$. 
We assume a solid angle of 10~msr and an expected momentum resolution of $\approx 
10^{-3}$. These properties are certainly satisfied by the spectrometers 
proposed as part of the Jefferson Lab energy upgrade. 
 
The pion rates have been estimated using the code EPC~\cite{EPC} and a 
parameterization of SLAC experimental data on  pion yields. The problem in these 
estimates lies in the fact that the kinematics measured at SLAC have 
very little overlap with those examined here. Similarly the 
parameterizations employed in the code EPC are not optimized for this 
kinematical region. We therefore use the yields obtained only as a 
rough guide. In a real experiment proposal these models would need 
to be refined.  Data from the approved $x>1$ measurement at 
$6$ GeV~\cite{e02019} will allow us to refine the pion and charge-symmetric 
background rate estimates. 
 
The count rates have been evaluated for an bin size of $\Delta x  = 
\pm 0.1$. The obtained rates for Deuterium are listed in 
Table~\ref{tab1} and the ones for Aluminum in Table~\ref{tab2}.  
The cross sections come from the calculations shown in Figs.~\ref{Fig.f2d} 
and~\ref{Fig.fescale}, (with multi-nucleon correlations included for the Aluminum). 
For these estimates, we have not included radiative effects.  
  
\begin{table}[htp] 
\begin{center} 
\begin{tabular}{cccc} 
$\theta_{e}$ &  $Q^2$(GeV/c)$^2$ & ${d\sigma\over d\Omega d\nu} [{ 
nb\over sr\cdot GeV/c}]$ & events/hour  \\ \hline 
5  & 1.06  & 1.24E+5 & 4.30E+7 \\ 
10 & 3.87  & 4.48E+1 & 4.65E+4 \\ 
20 &11.47  & 3.74E-2 & 6.53E+1 \\ 
30 &18.01  & 2.07E-3 & 3.35  \\ 
40 &22.5   & 4.37E-4 & 5.58E-1 \\ 
\hline 
\end{tabular} 
\end{center} 
\caption[]{Cross Sections and Count Rates for Deuterium, $x=1.5$, including correlations.} 
\label{tab1} 
\end{table}

\begin{table}[htp] 
\begin{center} 
\begin{tabular}{cccc} 
$\theta_{e}$ &  $Q^2$ (GeV/c)$^2$ & ${d\sigma\over d\Omega d\nu} [{ 
nb\over sr\cdot GeV/c}]$ & events/hour  \\ \hline 
5  & 1.06  &7.37E+6  & 1.60E+8 \\ 
10 & 3.87  &4.83E+3  & 3.13E+5 \\ 
20 &11.47  &2.06E+1  & 2.25E+3 \\ 
30 &18.01  &2.20     & 2.22E+2 \\ 
40 &22.5   &6.57E-1  & 5.08E+1 \\ 
\hline 
\end{tabular} 
\end{center} 
\caption[]{Cross Sections and Count Rates for Aluminum, $x=1.5$, including correlations.}  
\label{tab2} 
\end{table} 
 
The rate estimates indicate that for Deuterium the highest practical $Q^2$ 
value is about $Q^2 = 18$~(GeV/c)$^2$ while the count rate from Aluminum is still quite large 
at $Q^2 = 23$~(GeV/c)$^2$ with the experimental conditions 
described above.  If we use the cross section with only two-body SRCs included, 
the rate will be significantly reduced, but we should still be able to approach 
$Q^2 = 23$~(GeV/c)$^2$.  In all these cases we will reach $Q^2$ values where we are  
clearly dominated by the inelastic processes at $x>1$.

\subsection{Tagged Structure Functions}    
   
Understanding  the role of the quark-gluon degrees of freedom in the    
hadronic interaction is tied strongly to  understanding  the    
dynamics  responsible for modification of the quark-gluon structure   
of bound nucleons as compared to   
free nucleons. These dynamics at present are far from understood, however. 
Almost two decades after the discovery of the nuclear EMC effect~\cite{EMC1} 
and increasingly precise measurements~\cite{EMC2,EMC3,EMC3b,EMC4,EMC5}    
of the ratios of structure functions of nuclei and the deuteron,    
we still know only that this effect  requires the presence of {\it some}  
non-nucleonic degrees of  freedom in nuclei. No consensus has been reached  
on the origin  of these components.

The $x$-dependence of the effect, while non-trivial, is rather smooth    
and has the same basic shape for all nuclei, making it  easy to reproduce 
using   a wide range of models with very different underlying assumptions.   
The only additional constraint available to date comes from measurements 
of   the $A$-dependence of the sea distribution in Drell-Yan reactions\cite{DY},  
which   poses a problem for several types of models. 
The combination of the inclusive DIS and Drell-Yan experiments is still  
not sufficient to identify  unambiguously the origin of the EMC effect. 
Inclusive experiments at Jefferson Lab, after the upgrade,   will considerably improve  
our knowledge of the EMC effect by (i) measuring the  EMC effect for the  
lightest nuclei~\cite{e00101}, (ii) studying the isotopic dependence of  
the effect, and (iii) separating the different twists in the EMC effect. 
Though such experiments will be very important, they are unlikely to lead to  
an unambiguous interpretation of the EMC effect. 
New   experiments involving more kinematical variables accessible to    
accurate measurements are necessary to overcome  this rather unsatisfactory  
situation.

We propose that studying semi-inclusive processes involving a   deuteron target,  
\begin{equation}   
\label{eD}   
e + d \to e + N + X,    
\end{equation}  in which  
 a nucleon is detected in the target deuteron fragmentation    
region, may help to gain insight into the dynamics of the  
nuclear structure function modification~\cite{FS85,CL,CLS91,CL95,MSS97}.   
Further important 
information could be obtained by studying the production 
of $\Delta$-isobars and excited baryons in similar kinematics. 
Although we focus the discussion here on 
the simplest case of a deuteron  
target, clearly similar experiments using  heavier nuclei ($A=3$, $A=4$) would  
provide additional information.

A glimpse of the information that may be obtained from such    
experiments has already been provided by analyses of the experimental data on  
deep-inelastic neutrino  scattering off nuclei~\cite{Ber78,Efr80}.  
Even with poor statistics,  these  experiments have shown that structure functions,  
tagged by   protons produced in the backward  hemisphere,   
are different from those determined in inclusive scattering.   
   
The semi-inclusive experiments which we contemplate should 
be able to answer the following  questions:   
What is the smallest inter-nucleon  distance (or largest relative momentum)  
for which the two nucleons in the  deuteron  keep their identities?  
What new physics occurs when this  identity is lost? What is the     
signal that an explicit treatment of nuclear   quark and gluon degrees 
of freedom  is necessary? How much do the measured effective bound structure  
functions $F_{2}$ differ from those of free nucleons?  
Studying this difference will lead us to a better 
understanding of the dynamics behind  
nuclear binding effects and their  relation to QCD.   
   
The tool which allows us to control the relevant distances in the deuteron    
is the knowledge of the momentum of the tagged (backward) nucleon.   
The dependence of the semi-inclusive cross section  on this  
variable will test the  various assumptions regarding  
the lepton interaction with one nucleon while it is in a very 
close proximity to another nucleon. These very same assumptions  
are the ones that underlie the different models of the EMC effect. The study 
of the semi-inclusive reaction using  a deuteron target will permit important insights 
into the deepest nature of many-baryon physics. 
 
\subsubsection{Theoretical Predictions for the EMC effect} 
\label{sec:emcmodels} 
 
The {\bf first group} of models, referred to as binding models, makes  
the simplest assumption regarding the lepton-nucleon interaction in a nucleus, 
namely, that nucleons maintain their nucleonic character.  
In these models, the  nuclear EMC effect is caused by the effects of nuclear binding,  
and the inclusive structure function  data can be understood in terms of   
conventional nuclear degrees of freedom---nucleons and pions---responsible for 
nuclear binding~\cite{CL,AKV,AKVa,DT,ET,ANL,ANLa,HM,MEC,MECa,BT}.   
The nuclear cross section is then expressed as a    
convolution of the nuclear spectral function with  the structure function    
of a {\em free}  nucleon (Eqn.~(\ref{conv})).  
The off-mass-shell effect (if present at all) in these models constitutes a small 
part of the EMC effect, while the pion contribution is simply added to  
the contribution given by Eqn.~(\ref{conv}).  
 
The use of the standard, conventional meson-nucleon dynamics of nuclear  
physics is not able to explain both the nuclear deep-inelastic and the Drell-Yan 
data. One can come to this conclusion using the light-cone sum rules 
for the nuclear baryonic charge and the light-cone momentum~\cite{FS88}. 
Another approach~\cite{MillerSmith,MillerSmith2} is to use the Hugenholtz-van Hove 
theorem~\cite{HvH} which states that nuclear stability (vanishing of pressure) 
causes the energy of the single particle state at the Fermi surface to be  
$m_A/A\approx 0.99 m_N$. In light front language,  the 
vanishing pressure is achieved by setting $P^+=P^-=m_A$. 
Since $P^+=\int dk^+f_N(k^+)k^+$ and the Fermi momentum is a relatively small value 
 the probability $f_N(k^+)$ for a nucleon to have a given value of $k^+$ must be narrowly 
peaked about $k^+=0.99 m_N\approx m_N$. Thus, the effects of nuclear binding and Fermi  
motion play only a very limited role in the bound nucleon structure function. The   
resulting  function  must be very close to that of a free nucleon unless some quark-gluon  
effects are included. In this approach some non-standard explanation involving quark-gluon degrees  
of freedom is necessary. 
 
 Attempts to go beyond the simplest implementation of pion cloud 
effects have been made in Ref.~\cite{KOLTUN}, where recoil effects were taken into
account.  As a result it was  argued 
that the predicted pion excess based on conventional correlated nuclear 
theory is not in conflict with data on the Drell Yan process or with the 
nuclear longitudinal response function. (Note that the pion enhancement enters in these 
two processes in  a different way as the
integrands of the integral over the pion light cone fraction, $y_{\pi}$, have different 
$y_{\pi}$ dependence.)
What does appear to be ruled out, however, are RPA theories in which there 
are strong collective pion modes \cite{KOLTUN}. 
At the same time  the presence of pions on the level consistent 
with the Drell-Yan data does not allow us to reproduce the EMC effect
(in a way consistent with the energy momentum sum rule)  without 
introducing an ultrarelativistic mesonic component in the nucleus wave function.
  Each  meson's  light-cone fraction  in the wave function is  
$\leq 0.03$ but, taken together, the mesons carry $\sim 4\%$ of the light-cone fraction
of the nucleus.

The {\bf second group} of models represents  
efforts to model the EMC effect in terms of  {\em off-mass-shell} structure  
functions, which can be defined by taking selective non-relativistic or 
on-shell limits of the virtual photon--off-shell nucleon scattering 
amplitude~\cite{MST,KPW,GL}. For example, the structure function of a bound nucleon of 
four-momentum $p$ can depend on the variable $\gamma\cdot p-m$~\cite{HM}. 
Such terms should enter the  calculation of the structure function 
only in the form of multi-nucleon contact interactions, so these models 
represent a parameterization of a variety of dynamical multi-nucleon 
effects.   
A microscopic, quark-level mechanism which can lead to such a 
modification is provided by the quark-meson coupling models~\cite{qmc,qmc1,qmc2}, 
in which quarks confined in different nucleons interact via the exchange 
of scalar and vector mesons. 
Here the wave functions of bound nucleons are substantially 
different than those of free nucleons, even though the quarks are 
localized.

The {\bf third  group} of models  assume that the main phenomenon 
responsible for the  nuclear EMC effect is a 
modification of the bound state wave function of   
 individual nucleons. In these models, the presence of   
extra constituents in nuclei or clustering of partons from   
different  nucleons is neglected.    
Considerable modification of the bound nucleon structure   
functions at different ranges   
of $x$ are predicted by these models. Two characteristic approaches are    
the color screening model of suppression of point-like configurations in  
bound nucleons~\cite{FS88,FS85} and  models of quark   
delocalization or  dynamical rescaling~\cite{CLRR,JRR,CLOSE,NP,GP}. 
These models do not include possible effects of partial restoration 
of chiral symmetry which would result in modification of the pion 
cloud at short internucleon distances.

The {\em color screening} models start from the observation that the 
point-like  configurations (PLCs) in bound  nucleons are suppressed due to 
the  effects of color transparency~\cite{FS88,FS85,JM96}.  Based on the  
phenomenological success of the quark counting rules for   
$F_{2N}(x,Q^2)$ at large $x$  it is further assumed that three  
quark PLCs give the dominant  contribution at $x \geq 0.6$ and    
$Q^2 \geq 10$~GeV$^2$. The suppression of this component in a bound nucleon  
was assumed   to be a  main source of the EMC effect in inclusive deep-inelastic    
scattering in Ref.~\cite{FS85}. The size of the effect  was estimated to be  
proportional to the internal momentum of the target 
$k$ (to the virtuality of the interacting nucleon)~\cite{FS85}:   
\begin{equation}   
\delta_A({\bf k}^2) = (1+z)^{-2}, \ \ \ \ \    z = ({\bf k}^2/m_N + 
2\epsilon_A)/\Delta E_A, 
\label{plc}   
\end{equation}    
where $\Delta E_A = \langle E_i - E_N \rangle \approx m^* - m_N$, is    the 
energy denominator characteristic of  
baryonic excitations.  For $m^*=1.5-1.7$~GeV  the model reproduces reasonably the  
existing data on the EMC effect. Note that the considered  
suppression does not lead to a noticeable change  in the average 
characteristics of nucleons in nuclei~\cite{FS85}: for example, PLC suppression of  
Eqn.~(\ref{plc}) will lead to only $2\%$ suppression of the cross section for  high $Q^2$ 
scattering off a bound nucleon in quasi-free kinematics at $x=1$~\cite{FSS90}.

In {\em quark delocalization/rescaling} models it is assumed that    
gluon radiation occurs more efficiently in a nucleus than in a free nucleon    
(at the same $Q^2$) due to quark delocalization in two    
nearby nucleons. It is natural to expect that 
such a delocalization will grow with   
decreasing  inter-nucleon distance. Within this model the    
effective structure function of the nucleon can be written as:   
\begin{equation}   
F_{2N}^{\mathrm{eff}}(x,Q^2) = F_{2N}\left(x,Q^2\xi(Q^2,k)\right),    
\label{rsc}   
\end{equation}   
where  $\xi(Q^2)$ is estimated from the observed EMC effect in  
inclusive  deep inelastic cross sections, and its $Q^2$-dependence is  
taken from the generic form of the QCD evolution equations.  
 The $k$-dependence  in $\xi(Q^2,k)$ is modeled based on  
the assumption that the quark delocalization grows with increasing  
virtuality of a bound nucleon~\cite{MSS97}. 
    
In the above   classes of models the cross section for fast backward nucleon    
production in the reaction of Eqn.~(\ref{eD}) can be represented as    
follows~\cite{MSS97}:   
\begin{equation} 
{d\sigma^{e D \to e p X} \over 
 d\phi dx dQ^2 d(\log\alpha) d^2p_\perp }    
\approx  {2\alpha_{em}^2 \over x Q^4} (1-y) S(\alpha,p_\perp)    
    F^\mathrm{{eff}}_{2N}\left(\tilde x,\alpha,p_\perp,Q^2\right),   
\label{2N} 
\end{equation}  
where $S(\alpha,p_\perp)$ is the  
nucleon spectral function of the    
deuteron, and $F^{eff}_{2N}$ is the effective structure 
function of the bound nucleon, with $\tilde x \equiv x/(2-\alpha)$. 
The variable $\alpha= (E_s - p_{sz}) / m_N$   
is the light-cone momentum    
fraction of the backward nucleon with $p_{sz}$  negative for backward 
nucleons.   Both spectral and structure functions can be determined from  
the  particular models    discussed above~\cite{MSS97}.

An alternative scenario to the models discussed above is based on the idea discussed 
since the 1970's that two (three) nucleons coming sufficiently close together may form a kneaded
multiquark  state
\cite{sixquark1,sixquark2,sixquark3,PV,CH,gm6q,gm6q1,gm6q2,gm6q3,MULDERS1,MULDERS2}.
An example of such a state is a bag of six quarks. 
Multiquark cluster  models of the EMC effect were  developed in a number 
of papers~\cite{CH,LS,KS85,LS}).
In {\bf six-quark} ($6q$) models electromagnetic scattering 
from a $6q$ configuration is determined from a convolution of 
the structure function of the $6q$ system with the 
fragmentation functions of the five- (or more, in general) quark 
residuum\cite{CLS91}. These types of models cannot be represented through 
the convolution of a nucleon structure function and spectral function as in 
Eqn.~(\ref{2N}). 
Since the quarks in the residuum depend on the flavor of the struck 
quark, one finds\cite{CL95}: 
\begin{equation}   
{d\sigma^{e D \to e p X} \over    
 d\phi dx dQ^2 d(\log\alpha) d^2p_\perp }    
\approx {2\alpha_{em}^2 \over Q^4} (1-y) 
        \sum\limits_i x e_i^2 V_i^{(6)}(x)   
{\alpha\over 2-x}D_{N/5q}(z,p_\perp)\ , 
\label{6q}   
 \end{equation}   
where the sum is over quark flavors. 
Here $V_i^{(6)}$ is the distribution function for a valence quark 
in a $6q$ cluster, and $D_{N/5q}(z,p_\perp)$ is the fragmentation 
function for the $5q$ residuum, i.e., the probability per unit $z$ 
and $p_\perp$ for finding a nucleon coming from the $5q$ cluster. 
The argument $z$ is the fraction of the residuum's light-cone  
longitudinal momentum that goes into the nucleon: 
$  z = {\alpha\over 2-x}.   $

\subsubsection{Observables} 
 
\begin{figure}[ht]   
\vspace{-1cm}   
\begin{center}   
\epsfig{width=4.4in,file=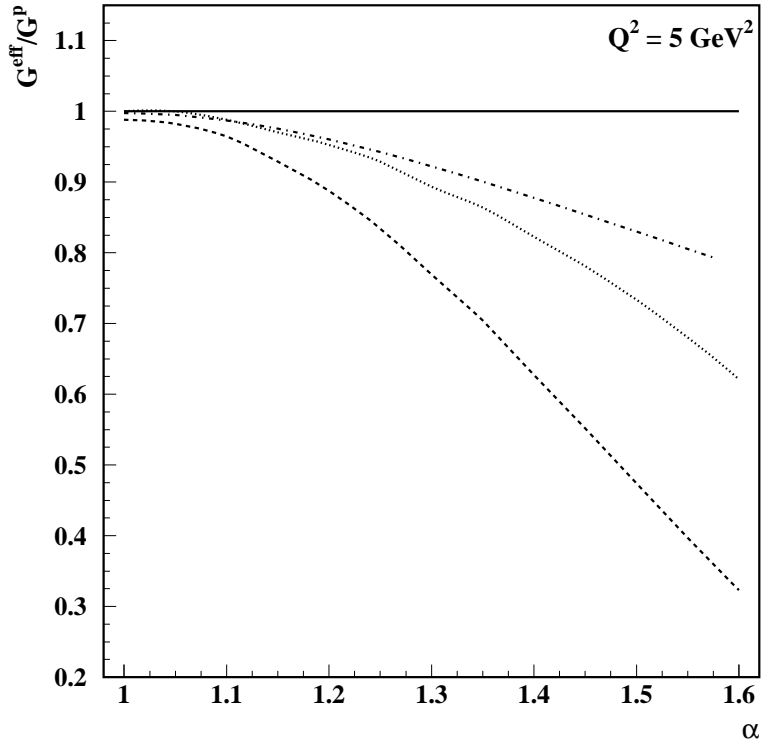}   
\end{center}   
\vspace{-1cm}   
\caption{The $\alpha$-dependence of $G(\alpha,p_\perp,x_1,x_2,Q^2)$,    
        with $x_1=x/(2-\alpha)=0.6$ and $x_2=x/(2-\alpha)=0.2$.    
        $G^{eff}(\alpha,p_\perp,x_1,x_2,Q^2)$ is normalized to    
        $G^{eff}(\alpha,p_\perp,x_1,x_2,Q^2)$   
        calculated with the free nucleon structure function with $p_\perp=0$.   
        The dashed line is the color screening model~\protect\cite{FS85}, dotted is the    
        color delocalization model~\protect\cite{CLRR}, and dot-dashed the off-shell  
        model~\protect\cite{MST}.}   
\label{Fig.Gratio} 
\end{figure}

Guided by the expectation~\cite{MSS97} that final state interactions 
should not strongly depend on $x$, Eqn.(\ref{2N}) suggests it is 
advantageous  to consider the ratio of cross sections  in two different 
bins of $\tilde{x}$: one where the EMC effect is small 
($\tilde{x}\sim 0.1-0.3$) and  one where the EMC effect is large 
($\tilde{x}\sim 0.5-0.6$)~\cite{FS88,FS85}. 
We suggest therefore measuring  the ratio $G$, defined as: 
\begin{eqnarray}    
G(\alpha,p_\perp,x_1,x_2,Q^2)&\equiv&    
\left.   
{ d\sigma (x_1,\alpha,p_\perp,Q^2) \over    
  dx dQ^2 d(\log\alpha) d^2p_\perp }    
\right/   
{ d\sigma (x_2,\alpha,p_\perp,Q^2) \over    
  dx dQ^2 d(\log\alpha) d^2p_\perp }    
\nonumber\\   
&=& {F^{eff}_{2N}(x_1/(2-\alpha),\alpha,p_\perp,Q^2) \over    
     F^{eff}_{2N}(x_2/(2-\alpha),\alpha,p_\perp,Q^2)}\ . 
\label{G}   
\end{eqnarray}   
Since the function $G$ is defined by the ratio of cross sections at the    
same $\alpha$ and $p_\perp$, any uncertainties in the spectral function    
cancel. 
This allows one to extend this ratio to larger values of $\alpha$,   
thereby increasing the utility of  semi-inclusive reactions. 
Figure~\ref{Fig.Gratio} shows the $\alpha$-dependence of   
$G(\alpha,p_\perp,x_1,x_2,Q^2)$ at $p_\perp=0$ for  several different models.   
The values of $x_1$ and $x_2$ are selected to fulfill the condition    
$x_1/(2-\alpha)=0.6$ (large EMC effect in inclusive measurements)    
and $x_2/(2-\alpha)=0.2$ (essentially no EMC effect). 
Note that while they yield similar inclusive 
DIS cross  sections, the models predict significantly different values of 
the ratio function $G$. 
   
\medskip 
 
{}From Eqn.~(\ref{2N}) one further observes that in all models that 
do not require the mixing of  quarks from different nucleons, 
the $x$-dependence of the cross section is confined (up to the kinematic 
factor $1/x$)  to the argument of the tagged nucleon structure 
function, $F_{2N}^{eff}$. 
On the other hand, for $6q$ models the $x$-dependence reflects the 
momentum distribution in the six-quark configuration, 
Eqn.~(\ref{6q}). 
As a result it is useful to consider the $x$-dependence of  the 
observable defined as: 
\begin{equation}   
R = \left.  {d\sigma\over dxdQ^2d\log(\alpha)d^2p_\perp}  \right/ 
{4\pi\alpha_{em}^2 (1-y)\over x Q^4}\ . 
\label{R}   
\end{equation} 
For the convolution-type models discussed above, the ratio $R$ is 
just the product of the spectral function, $S(\alpha,p_\perp)$, and the 
effective nucleon structure function, $F_{2N}^{eff}(\tilde x,\alpha,p_\perp,Q^2)$. 
In the on-shell limit the $x$-dependence of $R$ is therefore 
identical to that of the free nucleon structure function.  
 
\begin{figure}[ht]   
\vspace{-0.4cm}   
\begin{center}   
\epsfig{ 
width=4.4in,file=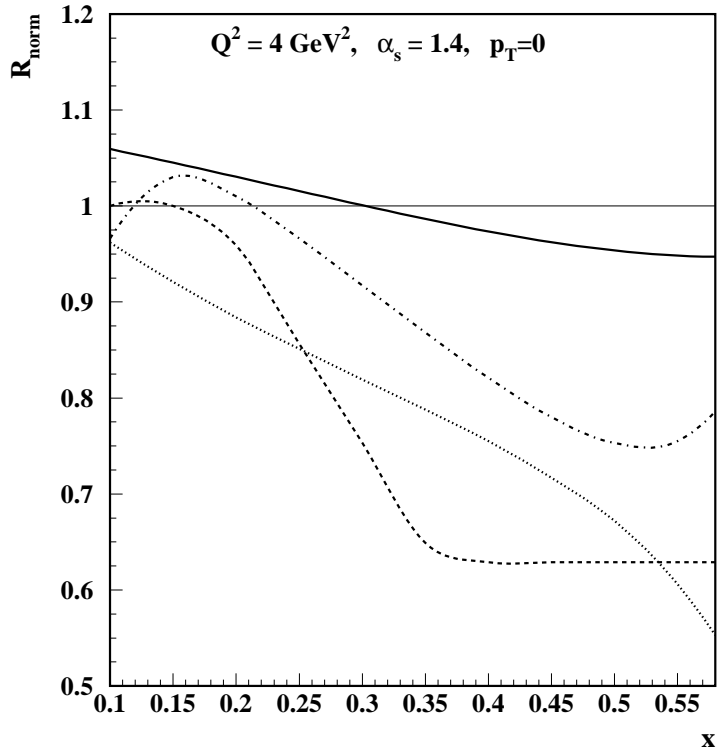}   
\end{center}   
\vspace{-1cm}   
\caption{The $x$-dependence of the normalized ratio $R$ defined in 
        Eqn.~(\ref{R}). The solid, dashed, dotted and dot-dashed curves 
        represent the predictions of off-shell (covariant spectator) 
        \protect\cite{MST}, color screening \protect\cite{FS85}, 
        quark delocalization \protect\cite{CLRR} and six-quark 
        \protect\cite{CL95} models, respectively. The binding models (group one) 
        predict the ratio of one - thin solid line.}   
\label{Fig.Rratio}   
\end{figure}

Figure~\ref{Fig.Rratio}  presents the $x$   
dependence of $R$ normalized by the same ratio calculated using  
Eqn.(\ref{2N}) with $F_{2N}^{free}$. The calculations are  performed 
for different models at $\alpha=1.4$  and $p_\perp=0$. 
Within the off-shell/covariant spectator model \cite{MST} $R$  exhibits 
a relatively weak dependence on $x$, while $6q$ models predict a 
rather distinct $x$-dependence.  Note that the models from  the first group  
(binding models)  does not produce any effect for this ratio. 
This ratio exposes a large divergence of predictions, all of 
which are obtained from models which yield similar EMC effects for 
inclusive reactions. 
Jefferson Lab at $11$ GeV should therefore have a unique potential 
to discriminate between different theoretical approaches to the 
EMC effect, and perhaps reveal the possible onset of the $6q$ 
component of the deuteron wave function. 
 
\begin{figure}[ht]   
\vspace{-0.4cm}   
\begin{center}   
\epsfig{width=4.0in,file=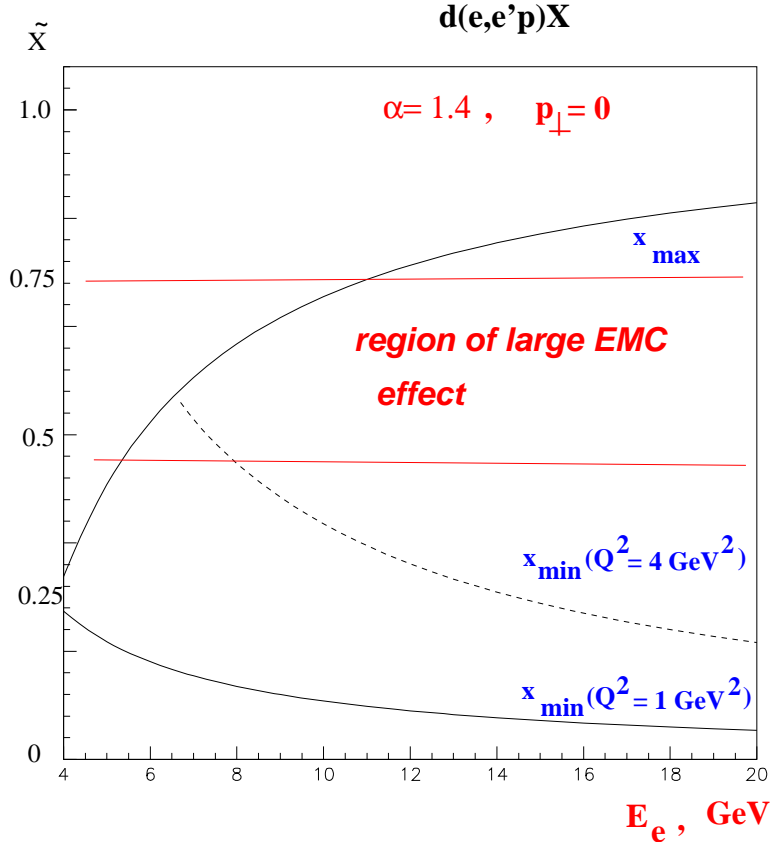}   
\end{center}   
\vspace{-0.4cm}   
\caption{The scaling window for $\alpha=1.4$.   
The upper curve is defined by the requirement that    
the mass of the produced final hadronic state $W\geq 2$~GeV.}   
\label{Fig.scalingregion}   
\end{figure}

An important advantage of the reactions considered with 
tagged nucleons, in contrast to inclusive reactions, is that any 
experimental result will be cross checked by the dependence of the 
cross section on $\alpha$ and $p_\perp$. 
Figure~\ref{Fig.scalingregion} shows the accessibility of  the 
scaling region as a function of incoming electron energy~\cite{CL95}. 
Values of $x$ between the curves labeled by $x_{\rm min}$ and $x_{\rm max}$ 
can be reached in the scaling region for a given incoming electron energy 
$E_e$.  

\subsubsection{Extraction  of the ``Free'' Neutron DIS Structure Function} 
 
Aside from providing a qualitatively new insight into  the origin of the 
nuclear EMC    
effect per se, the measurements of the tagged events may also be useful    
for extracting the free neutron structure    
function from deuteron data.   
By selecting only the slowest recoil protons in the target fragmentation   
region, one should be able to isolate the situation whereby the virtual   
photon scatters from a nearly on-shell neutron in the deuteron.   
In this way one may hope to extract $F_{2n}$ with a minimum of 
uncertainties arising from modeling 
nuclear effects in the deuteron.   
        
\begin{figure}[ht]   
\vspace{-0.4cm}   
\begin{center}   
\epsfig{width=4.0in,file=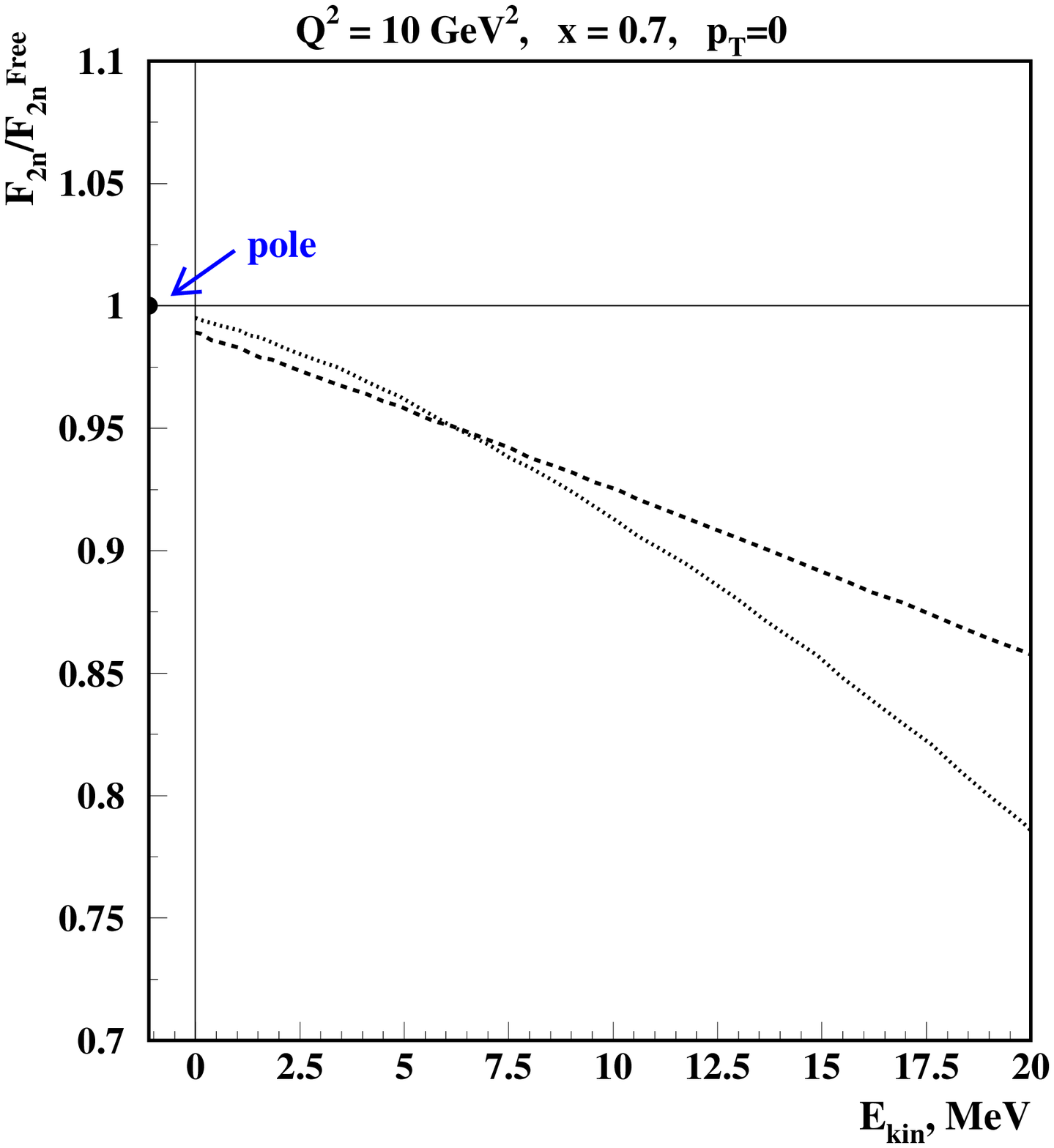}   
\end{center}   
\vspace{-1cm}   
\caption{The $E_{kin}$ dependence of the neutron structure function 
        extracted from $d(e,e'p_{\mathrm{backward}})X$ reactions within PWIA.  
        The effective structure function is normalized by the on-shell  
        neutron structure function. Dashed and dotted curves correspond    
        to the calculation within  Color Screening \protect\cite{FS85} 
        and Color Delocalization \protect\cite{CLRR} models, 
        respectively.} 
\label{Fig.taggedsf}   
\end{figure}   
   
One approach to extract the free $F_{2n}$ is to extrapolate the  
measured tagged neutron structure function  to the region of negative 
values of kinetic energy of the spectator proton\footnote{This method  
is analogous to the Chew--Low procedure for extraction of the cross  
section of scattering off a pion\cite{CL59}.}, where the pole 
of the off-shell neutron propagator in the PWIA amplitude is located  
($E_{kin}^{pole}=-{|\epsilon_{D}|-(m_{n}-m_{p})\over 2}$). 
The advantage of such an approach is that the scattering amplitudes 
containing final state interactions do not have singularities 
corresponding to on-shell neutron states. 
Thus, isolating the singularities through the extrapolation of 
effective structure functions into the negative spectator kinetic energy 
range  will suppress the FSI effects in the extraction of 
the free $F_{2n}$\cite{FSSprep}. 
Figure \ref{Fig.taggedsf} demonstrates that such an extrapolation can 
indeed be done with  the introduction of negligible   
systematic  errors.

\subsubsection{Experimental Objectives} 
 
\begin{figure}[ht] 
\center{\epsfig{width=4in,figure=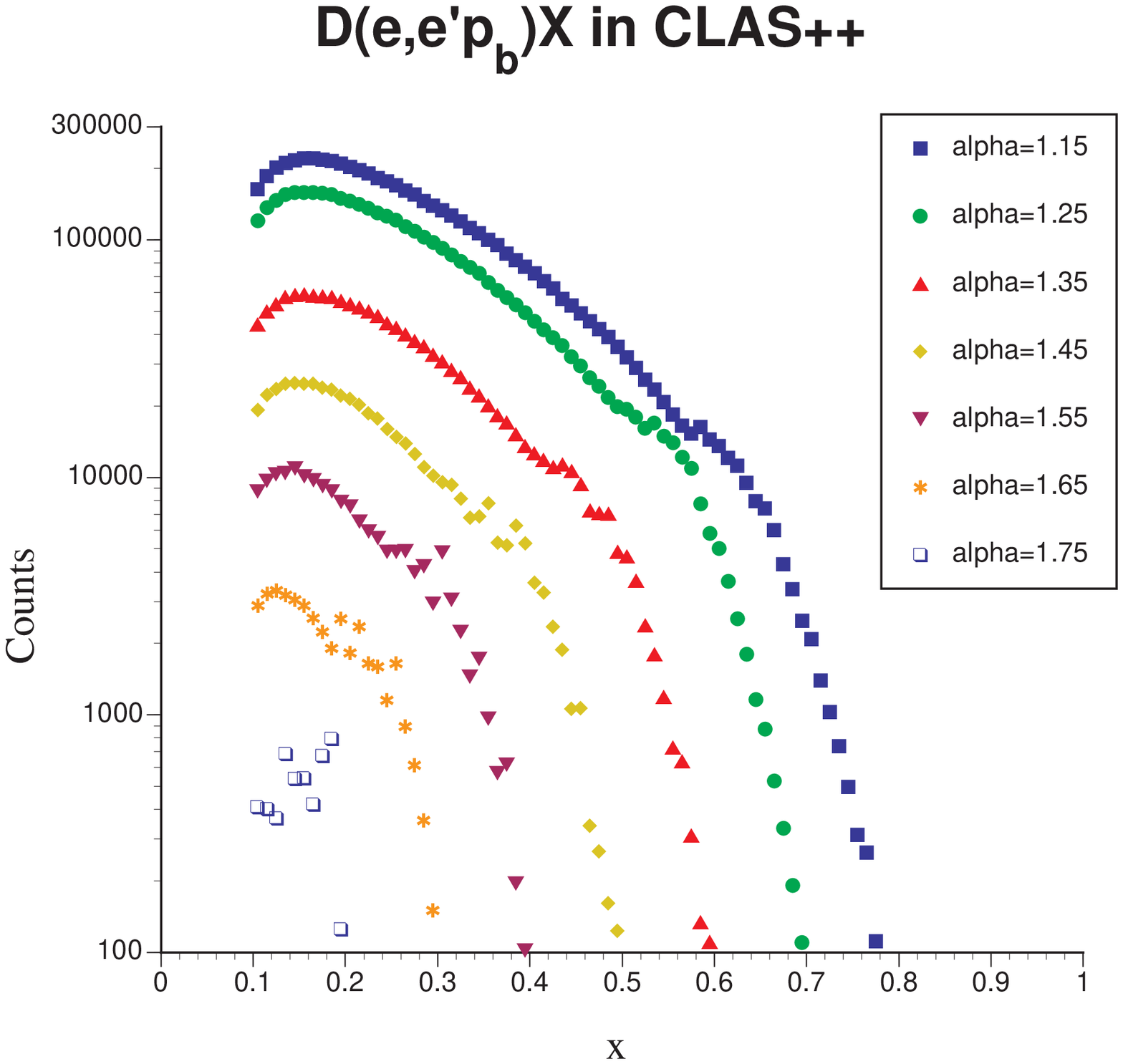}} 
\caption{ 
Kinematic coverage in Bjorken--$x$ and proton light-cone fraction 
$\alpha_S$ for the high--momentum part of the 
proposed experiment. The count rates have 
been estimated for a 20 day run with the standard CLAS++ configuration.  } 
\label{Deeps} 
\end{figure} 
 
As described in the previous section, our goal is to measure the 
reaction $d(e,e' p_{\mathrm{backward}})X$ over a large range in the electron variables 
($x$, $Q^2$) and the backward proton kinematics ($\alpha$, $p_\perp$). 
The proposed experiment would use the upgraded ``CLAS$^{++}$'' (CEBAF  
Large Acceptance Spectrometer) with an $11$~GeV beam to detect 
scattered electrons in coincidence with protons moving backward 
relative to the momentum transfer vector, $\bf q$. 
A large acceptance spectrometer is 
required since the proton is selected in spectator kinematics  
(small to moderate momentum up to $700$~MeV/c 
anywhere in the backward hemisphere) 
and is uncorrelated with the electron direction. 
Scattered electrons will be detected by the upgraded forward spectrometer 
with two sets of Cerenkov counters, time of flight counters, three tracking 
regions and pre-shower and total absorption electromagnetic calorimeters 
for highly efficient electron/pion separation. Depending on the momentum 
range of interest, two different detector/target arrangements will be used 
for the detection of the backward--moving proton. 
 
The first case involves the use of  a dedicated integrated target-detector 
system with a $5$ atmosphere deuterium gas cell as target ($30$~cm long 
and $0.5$~cm diameter) and a multilayer GEM (Gas Electron Multiplier) 
detector surrounding the target cell. By minimizing 
all materials in the path of large angle tracks, the threshold 
for detection of backward--moving protons can be lowered to about $70-80$~MeV/c. One 
expects that nucleon structure modifications and off-shell effects will 
be small at these momenta, and this method can be used to extract 
unambiguously the free neutron structure function $F_{2n}(x)$ up to  
very high values of $x$ ($\approx 0.8$).  
This measurement is of fundamental importance, since 
presently our knowledge of the neutron structure function  
at high $x$ is rather poor. 
At the same time, it will  supply the 
``low momentum`` part of the nucleon momentum dependence of the 
effective off--shell structure function, $F_{2n}^{eff}$, and thus serve 
as a baseline for the non-nucleonic effects  
which are expected at higher proton momenta. 
This target-detector system is presently under development 
and will be used for an exploratory measurement at $6$~GeV beam  
energy. Together with CLAS$^{++}$ and an $11$~GeV beam at a luminosity 
of $0.5\cdot 10^{34}$ cm$^{-2}$s$^{-1}$, a statistical 
precision of better than $\pm 5\%$ on $F_{2n}$  
out to the highest values of $x$ would be obtained 
with 40 days of data taking. 
 
In the second case, a central detector of CLAS$^{++}$, 
with superconducting solenoid, tracking and time-of-flight 
detectors would be used to measure backwards--going protons  
with momenta above $250$~MeV/c. 
With these higher momenta, one achieves great sensitivity to 
modifications of the neutron structure because of the 
proximity of the ``spectator'' proton. 
 
The dependence of the structure function  
$F_{2n}^{eff}(x/(2-\alpha), \alpha, p_\perp, Q^2)$  on the proton 
momentum from about $250$~MeV/c to over $600$~MeV/c can be extracted  
at fixed $x$ and $Q^2$. The experiment  will simultaneously cover  
a large range in $x$ and $Q^2$, allowing  detailed comparisons with 
the different models described in the previous section. 
Due to the higher momentum threshold, one can use a standard 
liquid deuterium target and the full CLAS$^{++}$ luminosity of 
10$^{35}$ cm$^{-2}$s$^{-1}$. 
 
Fig.~\ref{Deeps} shows estimates of  the 
expected number of counts for a $20$ day run 
as a function of $x$ for seven bins in the light-cone fraction $\alpha$ of 
the backward proton. One can clearly see the kinematic shift 
due to the motion of the struck neutron, which we can fully 
correct using the proton kinematics. 
It is  clear that  good statistics for a large range in $x$ and in   
$\alpha$ (the highest bin corresponds to more than 
$600$~ MeV/c momentum opposite to the direction of the $\bf q$ vector) 
would be obtained. Together with the low--momentum results, these data  
can be used to put the various models described in the previous section  
to a stringent test.

\section{Observation of Color Coherent Phenomena at  Intermediate Energies }  
 
\subsection{Basic aspects of color coherence and color transparency}   
 
\noindent 
QCD displays some of its special characteristics as a theory involving 
$SU(3)$-color by its prediction of novel effects in coherent 
processes. The basic idea is that the effects of gluons emitted by a  
color-singlet which is small-sized (or in a  point-like configuration) 
are canceled if the process is coherent. 
This is because, if the process is coherent, one computes the scattering  
amplitude by summing  the terms caused by gluon emissions from different quarks. 
The implication is that for certain hard exclusive processes, the effects of initial and/or 
final state interactions will be absent.

To observe color coherence effects it is necessary to find processes which are   
dominated by the scattering of  hadrons in a PLC and hence 
have amplitudes that can be  calculated using pQCD.   
A number of such processes have been suggested in the literature,   
and the corresponding QCD factorization theorems have been  proven for them.   
These processes include diffractive pion dissociation into two high transverse momentum    
jets~\cite{FMS93} and  exclusive production of  
vector mesons~\cite{Brod94,CFS97}.

Experiments at HERA which studied exclusive production of vector mesons in deep inelastic 
scattering (recently reviewed in~\cite{AC99}), have convincingly  confirmed   
the basic pQCD predictions -- a fast increase of the cross section
with energy at large $Q^2$, dominance of the  longitudinal  photon cross section and a weaker 
 $ t$-dependence of the $\rho$-meson production at large $Q^2$ relative to $J/\psi  $ photoproduction.

A distinctive feature of  processes such as  di--jet and vector meson   
production  is that, in the case of nuclear   targets, the incoming $q\bar q$ pair  
does not experience attenuation for the range of $x$ for which gluon shadowing  
is small ($x\geq 0.02$). This is the Color Transparency (CT)   phenomenon.   
As a result, the   amplitude  of the corresponding nuclear coherent processes  
at $t=0$,  and the cross section of  quasi-elastic processes are each 
proportional to  the nucleon number $A$, a result  which is    
vastly  different from the results usually obtained 
in processes involving soft hadrons. 
     
Color transparency, as predicted by pQCD, was directly observed in the Fermi Lab experiment
E791 which investigated  the exclusive  coherent production of two jets in the   
process $\pi +A \to 2 {\rm \ jets} + A$  at $E_{\pi}=500$~GeV.  
The observed $A$-dependence of the process~\cite{Ashery,Ashery2}  
is consistent with the predictions of~\cite{FMS93}, which lead   
to a  seven times larger platinum/carbon ratio   than soft physics would.    
The study of this reaction also allowed  measuring the pion $q\bar q$   
wave function~\cite{Ashery,Ashery2}, which turned out to be close to the   
asymptotic one at $k_\perp\geq 1.5$~GeV/c.   Evidence for  
color transparency effects was reported also in     
incoherent vector meson production in   DIS scattering of muons~\cite{E665}.

Hence we conclude that  the general concepts of CT in pQCD domain are now  
firmly  established for high energy processes:
the presence of PLCs in vector mesons and pions  and the  
form of the  small size $q\bar q$ dipole-nucleon    
interaction  at high energies 
are well  established experimentally\footnote{At sufficiently small $x (\leq 0.01)$ which 
can be achieved at RHIC, HERA,  and LHC,  the quantum field theory treatment of the CT predicts a 
gradual disappearance of the CT and the onset of the color opacity phenomenon.}

It is natural to apply the CT ideas to address the question of interplay of   
small  and large distances in various high-momentum transfer processes 
at intermediate energies. This involves three key elements:  
(i) the presence of configurations of small transverse size  in hadrons, 
(ii) the small size configurations not expanding as they exit the nucleus
(i.e. a  large coherence length) at high energies, which leads to the possibility  
of considering the small-sized configurations as frozen during the collision, and 
(iii) the weakness of the interaction of   small color singlet objects at high energies  
(for example, for a small  color $q\bar q$ dipole of transverse size $r_t$,   
the effective cross section is $\sigma_{\rm eff}\sim r_t^2$).  
It also is vital that the experiment be performed with sufficient precision to be  
certain that no additional pions be created. 
This is necessary to maintain the exclusive or nearly exclusive nature of the 
experiment, which is required for the necessary quantum interference    
effects to dominate the physics.     
The considered effect in general is analogous to the reduction of the  
electromagnetic interaction strength of the electrically neutral $Q^+Q^-$  
dipole  at small separations. However the uniqueness of QCD is in the  
prediction of the similar phenomena for color-neutral $qqq$ configurations. 
Establishing the existence of color coherence effects for three-quark systems 
would verify the color SU(3) nature of QCD, and remains an important unmet challenge.

\subsection{Goals For Intermediate Energy Studies} 
 
The  major   directions for study of CT related phenomena at   
Jefferson Lab are determined by the fact that one probes  
the transition  from soft to hard QCD regime. 
These studies include: 
 
\begin{itemize} 
\item Determining the interplay between the contribution of large and  
small distances for specific processes. 
\item  
Studying the interaction cross section of small objects in kinematic regions 
for which quark exchanges may play the dominant role as compared to  
gluon exchanges (at high energies the two-gluon exchange in $t$-channel dominates). 
\item Studying the dynamics of wave packet expansion. 
\end{itemize} 
 
All of these aspects can be addressed if one considers the fundamental 
physics involved with  
 the nucleon form factor at large $Q^2$ and the scattering amplitude for 
large angle hadron-nucleon elastic scattering.   
In QCD it is expected that at very large $Q^2$ the form factor and scattering 
amplitude are each  
dominated by  contributions arising  from   
the minimal Fock space components in the nucleon (hadron) wave function. 
Such components, involving the smallest possible number of constituents, are 
believed to be of a very small size, or to contain a PLC. 
To determine the values  of   $Q^2$ 
for which PLCs start to dominate,     
Brodsky~\cite{Brodsky82} and Mueller~\cite{Mueller82}     
suggested the  study of quasi-elastic hard    
reactions $l(h) +A \to l(h) + p +(A-1)^*$.    
If the energies and momentum transfers are large enough, one expects that 
the projectile and ejected nucleon travel through the nucleus as point-like 
(small size) configurations, resulting in a cross section proportional to 
$A$.

\subsection{Challenges for intermediate energy studies}   
   
To interpret  the physics of quasi-elastic reactions one has to address two  
questions:  
\begin{itemize} 
\item Can the  PLCs be treated  as a frozen during the time that the projectile  
  is passing through  the nucleus?   
\item  At what momentum transfer do the effects of PLCs dominate in the elementary  
 amplitude? 
\item Do they have small interaction cross sections? 
\end{itemize}  
These questions must be addressed if color transparency is to be studied at Jefferson Lab.  
If the momentum transfer is not large enough for the PLC to 
dominate or if the PLC is not frozen and  expands,   
then there will be strong final state interactions as the PLC 
moves through the nucleus.  
 
To appreciate the problems we shall discuss the effects of  
expansion in a bit more detail. Current color transparency experiments are  
performed in kinematic regions where the expansion of the produced small system  
is very important. In other words the length (the coherence length $l_c$) 
over which the PLC can move without the effects of time evolution 
changing the character of the wave function is too small, and this strongly  
suppresses   any effects of color transparency~\cite{FLFS88,FSZ90,jm90,jm90a,jm90b,jm90c,boffi}.    
The {\em maximal} value of $l_c$ for a given hadron $h$ can be estimated 
using the uncertainty principle: $l_c \sim {1\over \Delta M} {p_h\over m_h}$, 
in which $\Delta M$ is a characteristic excitation energy  
(for a small value of $m_h$ one can write $l_c\approx {2p_h\over \Delta M^2}$ 
with $\Delta M^2 =(m_{ex}^2-m_h^2)$ where  $m_{ex}^2$ is the invariant mass squared 
of the closest excited state with the  additive quantum numbers of $h$ (cf Eqn.~(\ref{eq2})). 
Numerical  estimates~\cite{FLFS88,jm90,jm90a,jm90b,jm90c} show that,  for the case of a nucleon  
ejectile,  coherence is completely lost at  
distances  $l>l_c \sim 0.3-0.5 {\rm\ fm} \times p_h$,   
where $p_h$ is measured in GeV/c.   
   
Two complementary  languages have been used to describe the effect of the loss of  
coherence. Ref.~\cite{FLFS88} used  the quark-gluon representation of the  
PLC wave function, to  argue that the  main effect is quantum diffusion of the  
wave packet so that     
\begin{equation}   
\sigma^{PLC,Q^2}(Z,Q^2) =(\sigma_{hard} + {Z\over l_c}[\sigma    
-\sigma_{hard}])   
\theta(l_c- Z) +\sigma\theta\left(Z-l_c\right),  
\label{eq:sigdif}   
\label{eq1}   
\end{equation}   
where $\sigma^{PLC}(Z,Q^{2})$  is the  effective total cross section 
for the PLC to interact  at a distance $Z$ from the hard interaction point. 
This equation is justified for the ``hard stage'' of time development   
in the leading logarithmic approximation when perturbative QCD can   
be applied~\cite{FLFS88,BM88,FS88,DKMT}.   
One can expect that Eqn.~(\ref{eq:sigdif}) smoothly  interpolates between the hard and  
soft regimes, at least for relatively large transverse sizes of the expanding  
system ($\geq 1/2$ of the average size) which give the largest contribution to   
absorption.     

The time development of the PLCs can also be  obtained by modeling the  
ejectile-nucleus interaction using a baryonic basis for  the PLC\cite{GFMS92}:    
\begin{eqnarray}   
\left| \Psi_{PLC}(t)\right> & = & \Sigma_{i=1}^{\infty} a_i \exp(iE_it)\left| \Psi_{i} \right>     
 \nonumber \\
& = & \exp(iE_1t)\Sigma_{i=1}^{\infty} a_i    
\exp\left({i(m_i^2-m_1^2)t\over 2p}\right)\left| \Psi_{i} \right>,   
\label{eq2} 
\end{eqnarray}   
where $\left| \Psi_{i} \right>$ are the eigenstates of the Hamiltonian with   
masses $m_i$, and $p$ is the momentum of the  
PLC which satisfies  $E_i \gg m_i$.     
As soon as the relative phases of the different hadronic components    
become large (of the order of one) the coherence is likely to be lost. 
It is interesting that numerically Eqns.~(\ref{eq1}) and~(\ref{eq2}) 
lead to similar results if a sufficient number of states is included in  
Eqn.~(\ref{eq2})~\cite{GFMS92}. It is worth emphasizing that, though both 
approaches model certain aspects of the dynamics of   the  expansion, a  complete 
treatment of this phenomenon in QCD  is absent.  

We next discuss the issue of the necessary momentum transfers required for 
PLCs to be prominent.   For electromagnetic reactions this question is related to  
the dominance of PLCs in the electromagnetic form factors of  
interacting hadrons. The later can be studied by considering the applicability 
of perturbative QCD in calculating the electromagnetic form factors. 
Current  analyses indicate that the leading twist approximation for   
the pion  form factor could become applicable (optimistically)  at    
$Q^2 \geq 10-20$~GeV$^2$; see e.g.  Ref.~\cite{JRS,JRSa}.  
For the nucleon case larger values of $Q^2$ may be necessary.   
However this does not preclude PLCs from being relevant for smaller values of 
$Q^2$. In fact, in a wide range of models of the  nucleon, such as constituent  
quark models with a singular  (gluon exchange type) short-range interaction or  
pion cloud models, configurations of  sizes substantially smaller 
than the average   one dominate in the form factor at 
$Q^2 \geq 3-4$~GeV$^2$; see  Ref.~\cite{fmsplc}. The message  
from QCD sum rule model calculations of the nucleon form factor is 
more ambiguous.

  For hadron-nuclear reactions the question of the dominance of PLCs is related  
to  theoretical expectations that  large angle hadron-hadron scattering is   
dominated by the hard scattering of  PLCs from each hadron. However this question  
is complex because one is concerned with placing larger numbers of quarks 
into a small volume.     Irregularities in the energy dependence of  
$pp$ scattering for $\theta_{c.m.}=90^\circ$, and large spin effects, have led 
to suggestions of the presence   of two interfering mechanisms in this  
process~\cite{rp88,bdet}, corresponding to    interactions of the nucleon  
in configurations of small and large sizes. See the review~\cite{pire}. 
 
The difficulties involved with using hadron beams in quasi-elastic reactions 
can be better appreciated by considering  a bit of history. 
The very first attempt to observe  color transparency effects was 
made at the AGS at BNL~\cite{Car88}.  The idea  was to see if  nuclei 
become transparent with an increase of momentum transfer in the 
$p+A\to p+p+(A-1)$ reaction.  As a measure of 
transparency, $T$, they measured the ratio:   
\begin{equation}   
T = {\sigma^{\mathrm{Exp}}\over \sigma^{\mathrm{PWIA}}},   
\label{T}   
\end{equation}   
where $\sigma^{\mathrm{Exp}}$ is the measured  cross section and  $\sigma^{\mathrm{PWIA}}$ is 
the    calculated cross section using plane wave impulse approximation 
(PWIA) when    no final state interaction is taken into account.    
Color transparency is indicated if $T$ grows   
and approaches unity with increasing energy transfer.

The experiment seems to support an increase of transparency  at 
incident proton momentum $p_{inc}=6 - 10$~GeV/c as compared to that   
at $E_p=1$~GeV;  see the discussion  in~\cite{FSZ94}.   
The magnitude of the effect can be  easily   described in color transparency  
models which include the expansion effect.   The surprising result of the  
experiment was that with further increase of momentum, ($\geq 11$~GeV/c), $T$ decreases.  
The first data from a new $(p,2p)$ experiment, EVA\cite{BNL98},  at $p_{inc}=6 - 7.5$~GeV/c 
confirmed  the findings of the   first experiment~\cite{Car88} and more recently      
EVA has reported measurements in a wider momentum range up to $14$~GeV/c. 
The data appear to confirm both the increase of   
transparency between $6$ and $9$ GeV/c and a drop of transparency at $12$ 
and $14$~GeV/c~\cite{BNL01}. 
 
The  drop in the transparency can be understood as a  
peculiarity of the elementary high momentum transfer $pp$ scattering amplitude,  
which contains  an interplay  of contributions of PLCs and large size 
configurations as suggested in~\cite{rp88,bdet}. A description of the drop in transparency based on these ideas was presented
in Ref.~\cite{jm94}.  However, it is evident that the interpretation of any  
experiment would be simplified by using an electron beam.   
   
The most general way to deal with each  of the challenges mentioned here is to  
perform relevant experiments using electrons at the highest possible values of  
energy and momentum transfer.

\subsection{Color Transparency in (e,e'N) and (e,e'NN) Reactions }    
  
\begin{figure}[ht]   
\begin{center}   
\epsfig{width=4.8in,file=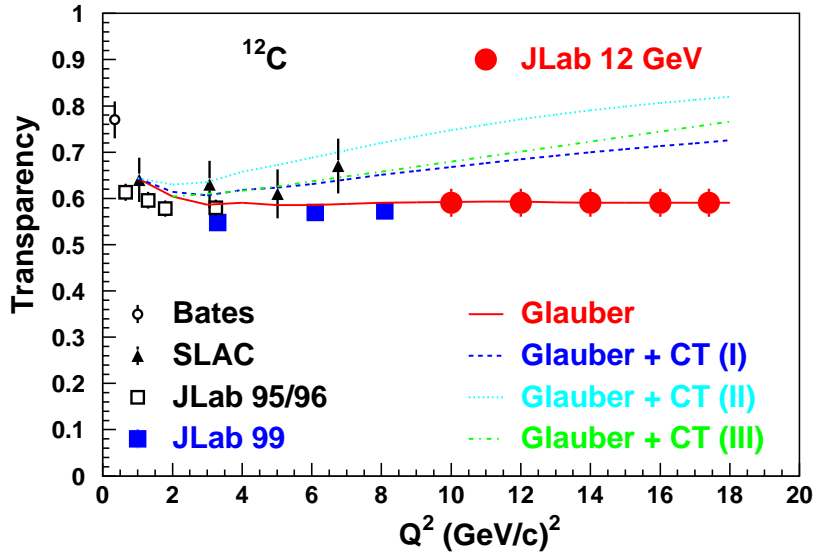}   
\end{center}   
\vspace{-0.9cm}   
\caption{\protect The $Q^2$-dependence of $T$ as defined in Eqn.~(\ref{T}).  
``Glauber'' - calculation within  
Glauber approximation, ``Glauber+CT(I) and Glauber+CT(II)'' - calculations  
including CT effects with expansion parameter $\Delta M^2=1.1$~GeV$^2$ and  
$\Delta M^2=0.7$~GeV$^2$ respectively~\cite{FMSS}, ``Glauber+CT(III)'' -   
CT effects are included according to Ref.~\cite{NSSW}.}   
\label{Fig.transparent}     
\end{figure}

The first step is to measure  a transparency similar to that of Eqn.~(\ref{T}) using 
electroproduction reactions. The first electron $A(e,e'p)$ experiment    
looking for color transparency was NE-18  performed at    SLAC~\cite{NE18,NE18a}.  
The maximum  $Q^2$ in this experiment is $\approx 7$~GeV$^2$,  which corresponds to  
$l_c \leq 2$ fm.    
For  these kinematics, color transparency models which included expansion effects    
predicted a  rather small increase of the transparency; see for example~\cite{FS88}.   
This prediction is consistent with the NE-18 data. However these data are   
not sufficiently accurate  either to confirm or to rule out color    
transparency on the level predicted by realistic color    
transparency models.  Recent Jefferson Lab experiments~\cite{quench,garrow} have been 
performed up to $Q^2=8$~GeV$^2$, and no effects of color transparency have 
been observed (see Fig.~\ref{Fig.transparent}). However, models of color transparency  
which predict  noticeable effects in the $(p,pp)$ reaction include versions which can also lead  
to almost no effects in electron scattering. In those models,  the effects of  
expansion are strong for the lower energy final state wave functions, but do 
allow some color transparency to occur for the initial state wave function. 
At $Q^2=8$~GeV$^2$, the momentum of the  proton ejected in electron scattering 
is about $5$~GeV/c,  which is still lower than the lowest momentum, $6$~GeV/c used at BNL.  
One needs to achieve a $Q^2$ of about $12$~GeV$^2$ to reach  a nucleon momentum for which BNL  
experiment observed an increase of the transparency. 
 
\begin{figure}[ht]   
\begin{center}   
\epsfig{width=4.4in,file=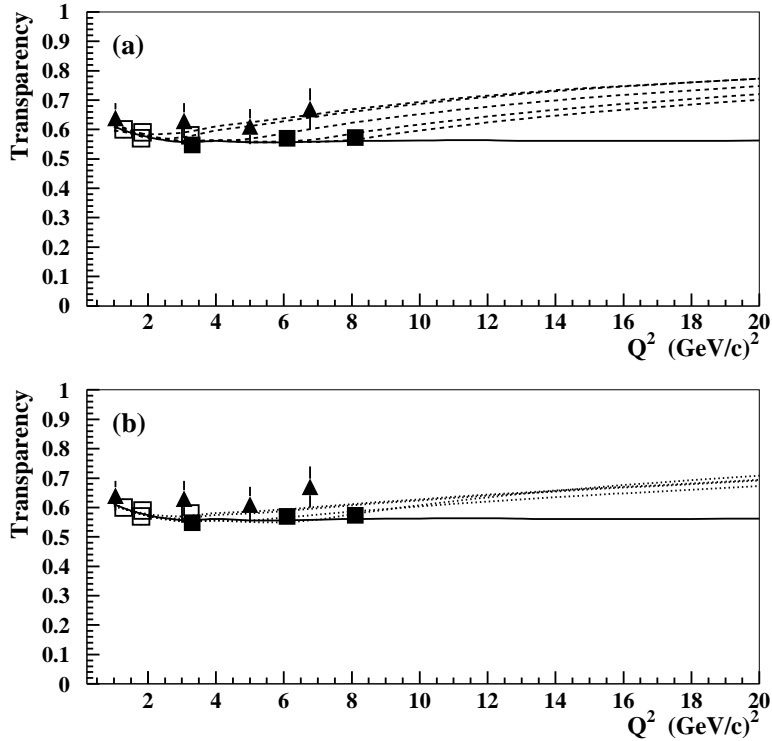}   
\end{center}   
\vspace{-0.9cm}   
\caption{The $Q^2$-dependence of T. The solid line is the prediction of  
the Glauber approximation. In (a) dashed curves correspond to the  
CT prediction with $\Delta M^2=0.7$~GeV$^2$ and with $Q^2_0=1$(upper curve),  
$2$, $4$, $6$ and $8$~GeV$^2$(lower curve). 
In (b) dotted curves correspond to the  
CT prediction with $\Delta M^2=1.1$~GeV$^2$ and with $Q^2_0=1$(upper curve),  
$2$, $4$ and $6$~GeV$^2$(lower curve). All calculations are normalized to  
the data at $Q^2=2$~GeV$^2$. The data are the same as in Fig.~\ref{Fig.transparent}. } 
\label{Fig.trans12lim}     
\end{figure}

The recent Jefferson Lab data~\cite{garrow} allow us to put some constraints on  the  
parameters defining the onset of CT. In Fig.~\ref{Fig.trans12lim} we analyze the lower  
limit of $Q^2_0$ at which PLCs are selected in $\gamma^*N$ scattering. Since our interest is  
only in the energy dependence of the transparency, we normalized the calculations to the  
data at $Q^2=2$~GeV$^2$ to avoid uncertainties related to the $Q^2$ dependence of quenching
\footnote{The transparency is defined in Eqn \ref{T} is inversely proportional to the nuclear quenching.
Hence a decrease of the quenching effect with an increase of $Q^2$ may mask the CT effects at 
intermediate $Q^2\leq 2 GeV^2$.}
The analysis is done  within the quantum diffusion model of CT for the range  of  
the expansion parameter $\Delta M^2$ 
($\Delta M^2=0.7$~GeV$^2$ in Fig.~\ref{Fig.trans12lim}(a) and $\Delta M^2=1.1$~GeV$^2$  
in   Fig.\ref{Fig.trans12lim}(b)) consistent with the EVA data.  
In the case of the slower expansion rate, Fig.~\ref{Fig.trans12lim}(a), the transparency  
is rather sensitive to $Q^2_0$ and the analysis yields a lower limit of $Q_0^2\approx 6$ GeV$^2$. 
For a faster expansion rate, Fig.~\ref{Fig.trans12lim}(b), the nuclear transparency  
is less sensitive to  $Q^2_0$, since for intermediate range of $Q^2$ the 
PLC expands  
well before it escapes the nucleus.  
The analysis in Fig.~\ref{Fig.trans12lim}(b) yields $Q^2_0\ge 4$~GeV$^2$.  
Combining these two analyses one can set the lower limit for the formation of PLCs at  
$Q^2_0 \approx 4$~GeV$^2$  (see  Eqn.~(\ref{range}) for implication of this limit 
in SRC studies).

The  upgrade of Jefferson Lab would improve the situation, by pushing the  
measurement of $T$ to a high enough $Q^2$   where the color 
transparency    predictions appreciably diverge from the predictions of  
conventional calculations (see Figs.~\ref{Fig.transparent} and~\ref{Fig.trans12lim}).    
Indeed,  the EVA data have established in a  
model independent way that  at least for nucleon momenta $\geq 7.5$~GeV/c,  
expansion effects are not large enough to mask the increase of the transparency.  
Hence measurements at $Q^2 \geq 14$~GeV$^2$, corresponding to comparable momenta of  
the  ejectile nucleon, would  unambiguously answer the question whether   
nucleon form factors at these $Q^2$ are dominated by small or large   
size configurations.  
 
\begin{figure}[ht] 
\begin{center} 
 \epsfig{width=4.8in,file=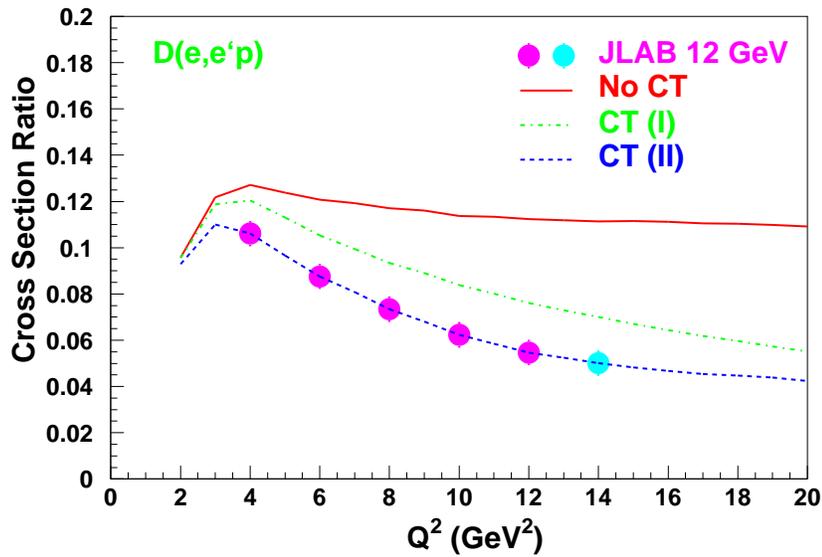}   
\end{center} 
\vspace{-0.9cm} 
\caption{The ratio of the cross section at $400$~MeV/c missing momentum to 
the cross section at $200$~MeV/c as a function of $Q^2$.  
Solid line corresponds to the GEA prediction. Dashed and dash-dotted lines
represent the quantum diffusion model of CT with $\Delta M^2=0.7$ and $1.1$~GeV$^2$ 
respectively. The drop with $Q^2$ in the color transparency models comes from a 
reduction in the rescattering of the struck nucleon, which is the dominant 
source of events with $P_m > k_F$.} 
\label{Fig.CT_ratio} 
\end{figure} 
   
Although $(e,e'N)$ measurements will allow an unambiguous check of    
the existence of color transparency, it will require a challengingly high   
accuracy  from experiments to  investigate the details of the expansion  
effects (Fig.~\ref{Fig.transparent}). Thus, although this is the simplest 
reaction to measure,  a much wider range of reactions would be necessary to 
build a sufficiently  complete picture of phenomenon and to scan the 
expansion of the small size  wave packets.

To obtain a  more detailed knowledge of the interaction of PLCs with   
nuclei we can also select a  processes  in which the ejectile could  
interact a second time during its propagation through   
the nucleus~\cite{FS91,double,FGMSS95} (double scattering reactions).  
This can be done by studying recoil nucleons with perpendicular (vs $\vec q$)  
momenta  $p_{s,\perp} \geq 200$~MeV/c. At low $Q^2$, the majority of such high 
momentum nucleons come from rescattering with the spectator nucleon in the 
nucleus. Therefore, the  number of such  
nucleons should decrease substantially with the onset  of CT which reduces  
the probability of rescattering.   
An important advantage of a double scattering reaction is that the 
disappearance of the   final state interactions can be studied using the 
lightest nuclei (D,${^3}$He,${^4}$He), for which wave functions are known 
much better and where one can use a generalized eikonal approximation, which  
accounts for the nonzero values of the momenta of recoil nucleons  
~\cite{FGMSS95,FSS97}
\footnote{Note that in conventional Glauber approximation,  
the momenta of recoil nucleons are neglected.}.  
Another advantage of double scattering reactions  is that inter-nucleon distances  
probed are not large, $1-2$~fm.  These distances are comparable to the coherence  
length for values of $Q^2$ as low as about $4-6$~GeV$^2$, and   
may provide evidence for a number of color coherent phenomena in  
the transitional $Q^2$  region. Ultimately, double scattering measurements  
will allow us to determine whether the lack of the CT  in $A(e,e'p)$  
reactions at $Q^2\geq 8$~GeV$^2$ region is  related to the large expansion 
rate of PLCs, or if it is because PLCs are not produced at all for these 
values of $Q^2$.

An appropriate measure for color  transparency in double scattering reactions  
is a ratio of  cross sections, measured at kinematics for which  
double scattering is dominant, to the cross section measured at 
kinematics where the effect of Glauber screening is more 
important.   Theoretical investigations of these reactions~\cite{double,FGMSS95}  
demonstrated that it is possible  to separate these two kinematic regions  
by choosing two momentum intervals for the recoil nucleon:  ($300-500$~MeV/c)  
for double scattering, and ($0-200$~MeV/c) for Glauber screening.   To enhance  
the effect of the final state interaction in both regions, the parameter $\alpha$, 
characterizing the light cone momentum fraction of the nucleus carried by  
the recoiling nucleon should be close to one ($\alpha = (E_s-p^z_s)/m \approx 1$,  
where $E_s$ and ${\bf p_s}$ are the energy and momentum of recoil 
nucleon in the final state). 
Thus, the suggested experiment will measure  the  $Q^2$-dependence of the  
following typical ratio at $\alpha=1$:     
 \begin{equation}   
R = {\sigma(p_s = 400~MeV/c)\over \sigma(p_s=200~MeV/c)}   
\label{rdouble}   
\end{equation}   
   
Figure~\ref{Fig.CT_ratio} shows this ratio, calculated within the generalized eikonal 
approximation~(solid line),  
and using the quantum diffusion model of CT with upper and lower values of the expansion 
parameter  $\Delta M^2$. 
   
It is worth noting that in addition to the $d(e,e'pn)$ process, one can consider  
excitation of baryon resonances  produced in the spectator kinematics,  
like $d(e,e'p)N^*$ and $d(e,e'N)\Delta$.   The latter process is of special   
interest for looking for  the effects of so-called chiral    transparency---the  
disappearance of the pion field of the ejectile~\cite{fms,chiral}.

\subsubsection{Experimental Objectives} 
 
The $A(e,e^\prime p)$ and $d(e,e^\prime p)$ experiments described in the 
previous Section are rather straightforward: they require a high-luminosity 
electron beam to access the very small cross sections at 
high-$Q^2$ and a set of two medium-resolution magnetic spectrometers to 
determine, with reasonable precision, the recoil nucleon's momentum and 
the nucleon's binding energy. 
 
In the case of the $A(e,e^\prime p)$ transparency measurements, the nuclear recoil 
momentum is typically restricted to a momentum 
smaller than the Fermi momentum $k_F$ ($\approx$ 250--300 MeV/c). 
The missing energy (identical to the binding energy plus nuclear excitation 
energy) is restricted to be well below pion production threshold 
($\approx$ 100 MeV). SRCs within a nuclear system push a sizable amount 
of the individual nucleons to large momenta and binding energies. As 
precise, quantitative evidence of this effect remains elusive, it is 
preferable to restrict oneself to the region of single particle strength 
described by above cuts in recoil nucleon momentum and missing energy. 
 
Using two medium-resolution magnetic spectrometers, 
with momentum and angular resolutions of order 0.1\% and 1 mr, one can 
easily make well-defined cuts in recoil nucleon momentum and missing energy. 
In the case of the $d(e,e^\prime p)$ cross section ratio measurement, a 
good determination of the recoil nucleon's momentum is a fundamental concern: 
the recoil nucleon momentum distribution drops steeply with missing 
momentum, even though the double rescattering mechanism partly counteracts 
this effect. This is illustrated by the absolute value of the cross section 
ratio in Fig.~\ref{Fig.CT_ratio}, $\approx$ 0.1, for recoil nucleon momenta of 400 MeV/c 
with respect to 200 MeV/c. Moreover, at 200 MeV/c the cross section 
varies with recoil nucleon momentum by about 30\% per 10 MeV/c. 
Hence, an absolute comparison of the measured $d(e,e^\prime p)$ cross 
section ratio with calculations requires determination of the recoil 
nucleon momentum value with precision much smaller than 10 MeV/c. 
 
The use of large non-magnetic devices, such as electromagnetic calorimeters, 
may be possible in limited cases, but will seriously affect the required 
missing momentum determination. Thus, for the $A(e,e^\prime p)$ and 
$d(e,e^\prime p)$ examples shown in Figs.~\ref{Fig.transparent}  and~\ref{Fig.CT_ratio},  
two magnetic spectrometers 
operating at a luminosity of 1$\times$10$^{38}$ electron-atoms/cm$^2$/s, with 
3 and 6 msr solid angle, respectively, were assumed. The larger solid angle 
magnetic spectrometer would detect the quasi-elastically scattered electrons, 
the smaller solid-angle would map the (relativistically boosted) Fermi cone. 
The latter magnetic spectrometer would need a momentum range between $4$~GeV/c 
($Q^2 \approx 6$ (GeV/c)$^2$) and 10 GeV/c ($Q^2$ $\approx$ 17 (GeV/c)$^2$). 
Missing energy and recoil nucleon momentum resolutions would still be better 
than 10 MeV and 10 MeV/c, respectively. 
 
Lastly, the estimated beam time for the projected uncertainties in 
Figs.~\ref{Fig.transparent}  
and~\ref{Fig.CT_ratio} would be less than one month of beam time for the 
$^{12}$C$(e,e^\prime p)$ transparency measurements, and one month of beam time 
for the d$(e,e^\prime p)$ cross section ratio measurements up to $Q^2$ = 
12 (GeV/c)$^2$, with one additional month required to push these ratio 
measurements to $Q^2$ = 14 (GeV/c)$^2$. For the cross section ratio 
measurements, one would need to determine the cross section yields at 
recoil nucleon's momenta of 200 and 400 MeV/s with two separate angle 
settings of the magnetic spectrometer (included in the estimated beam times). 
Note that the $d(e,e^\prime p)$ measurements will build on the 
existing Jefferson Lab  experiments\cite{KG,EGS} at $6$~GeV which plan to 
study the ratio $R$~(Eq.(\ref{rdouble})) up to $Q^2$ of $6$~GeV$^2$.

\subsection{Color Coherent Effects with Coherent Vector Meson Production 
off the Deuterium}  
 
Although the main emphasis in color transparency studies is given to the  
experiments with nucleon electroproduction, it is widely expected that one 
should  observe the onset of color coherence in meson electroproduction  at lower  
values of $Q^2$ than for the case of nucleon knockout.  It should be easier 
to find the quark and anti-quark of a meson close together to form a 
point like configuration, than to find the three quarks of a  
nucleon together. 
 
The QCD factorization theorem\cite{CFS97} for the exclusive meson 
production by the longitudinally polarized virtual photons demonstrates that
the $q\bar q$ PLCs dominate in the Bjorken limit and that
CT should occur both for the coherent and incoherent channels. In the
leading twist the exclusive meson production can proceed  through the
quark-antiquark and, in case of vector mesons, also through the two gluon ladder
exchange in t-channel\cite{CFS97}. 
At small $x$ (probed for example at HERA\cite{AC99}) the two gluon ladder exchange 
dominates in the production of $\rho$ and $\omega$-mesons. 
However at Jefferson Lab kinematics, production of $\rho$ and $\omega$-mesons is dominated by 
the quark exchange \cite{CFS97,GV}. The latter is confirmed by the 
analysis of HERMES data on $\gamma^* + p \rightarrow \rho+ p$  reaction\cite{GV}.
Additionally, the analyses of Ref.~\cite{GV,FKS96,Lech} indicate that the
leading twist approximation overestimates strongly (by a factor $\sim 4 $
for $Q^2\sim$ 4 GeV$^2$) the $\rho$-meson production cross
section. The suppression factor in leading approximation 
contribution is explained in Refs.\cite{FKS96,GV} as a higher twist effect
due to the finite transverse size of the photon 
wave function in the convolution integral involving the interaction block, 
virtual photon, and meson wave functions.
However, the model analysis of \cite{FKS96} indicates that the higher twist
effects may not interfere with the transverse localization of the vector meson
wave function leading to a possibility of the sizable CT effects already at 
$Q^2\ge $4 GeV$^2$. The suppression of the leading twist contribution was 
observed by \cite{GV} for the quark exchange channel for  
both vector meson and pion production. However the corresponding analysis of the 
transverse interquark distances is not yet available.
 
Studies of the CT for  meson production with the Jefferson  Lab  upgrade will be very
important for understanding of the onset of the leading twist contribution
and determining what transverse separations are important in the higher
twist contributions. An observation of  CT for  meson production
would allow us to use these processes at pre--asymptotic $Q^2$ for measuring
the ratios of different nucleon generalized parton distributions. In the
case of the vector meson production it is feasible to look for  CT both
in  coherent and incoherent scattering off
nuclei\cite{BM88,FS88,KNNZ,Brod94}. 
Studies of incoherent reactions require a
very good energy resolution in the mass of the residual system to suppress
 processes where a meson is produced in the elementary reaction,
processes like $\gamma_L +N \to M + \Delta$, as well as in a multi-step
processes like $\gamma_L +N \to \rho +N,
\rho+N^*\to \pi +N^{**}$. The first experiments looking for  CT effects in the
incoherent production of pions and $\rho$-mesons  were recently
approved at Jefferson Lab \cite{e01107,e02110}. 

Here we will focus on the reactions of the
coherent meson production in which  the background processes mentioned above 
are suppressed.  The first experiment dedicated to the studies of coherent 
production of vector mesons from nuclei at $6$~GeV approved recently at 
Jefferson Lab\cite{SKKpro}.
The proposed upgrade of Jefferson Lab, 
which will provide high energy, high intensity,  
and high duty factor beams makes systematic studies of 
these reactions a very promising area of study. 
 
\begin{figure}[th] 
\vspace{-0.4cm}   
\begin{center}   
\epsfig{width=4.8in,file=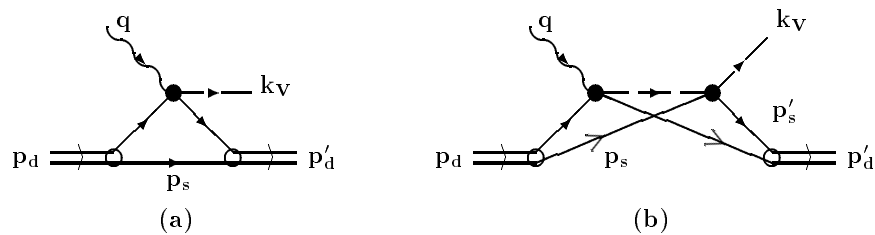}   
\end{center}   
\vspace{-0.4cm} 
\caption{Diagrams corresponding to single (a) and double scattering 
contributions in coherent vector meson electroproduction.}  
\label{Fig.edv} 
\end{figure} 
 
\begin{figure}[ht] 
\begin{center} 
\epsfig{width=4.0in,file=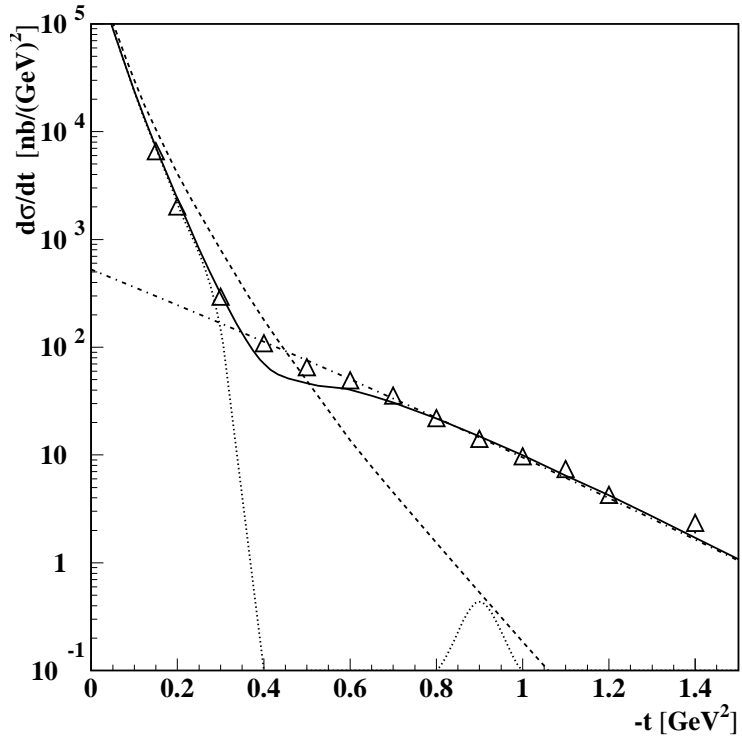}   
\end{center}   
\caption{Cross section of the coherent $\rho^0$ photo-production off 
deuteron. Data are from~\protect{\cite{slac71,overman}}. Dashed, dash-dotted, dotted 
and solid curves represent single scattering, double scattering, interference between 
single and double scattering, and full contributions respectively. Theoretical 
predictions are based on the vector meson  dominance (VMD) 
model~\protect{\cite{FPSS98}}.}  
\label{rhoexp} 
\end{figure}

The most promising channel for studying color coherent effects with  
meson electroproduction is the coherent production of vector mesons off 
deuteron targets: 
\begin{equation} 
e+d \to e' + V + d' 
\label{vmpro} 
\end{equation} 
where ``$V$'' is the $\rho$, $\omega$ or $\phi$ meson. This reaction is a 
unique channel for studying CT for the following reasons: 
\begin{itemize} 
\item Due to the large {\it photon-vector~meson} coupling the cross section of the 
process is large, and at high energies and small $Q^2$($< 1$~GeV$^2$) is well 
understood  in the framework of the vector meson dominance (VMD) model; 
 
\item The deuteron is the theoretically best understood nucleus. It has zero isospin 
and as a result, in the coherent channel, $\rho-\omega$ mixing will be 
strongly suppressed, since  
the $\rho$ and $\omega$ have isospin one and zero respectively. 
The technical advantage of using the deuteron in coherent reactions is the 
possibility of detecting the recoil deuterons. 
 
\item  Coherent production of vector mesons off deuterium is characterized by two 
contributions:  
single and double scattering contributions (Fig.~\ref{Fig.edv}). Moreover, it is a 
well known fact from the photo-production experiments~\cite{slac71,overman} that at 
large  $-t\geq 0.6$ GeV$^2$,  the double scattering contribution can be unambiguously 
isolated  (Fig.~\ref{rhoexp}). 
\end{itemize} 
 
The strategy of CT studies in the coherent reaction of Eqn.~(\ref{vmpro}) is somewhat 
similar to the strategies of studying CT in double scattering $(e,e'NN)$ reactions. 
First one has to  
identify kinematics in which double scattering effects  can be isolated from  
Glauber type screening effects (corresponding to the interference term of single and  
double scattering amplitudes).  The availability of  the  $t$-dependence of the 
differential cross section allows us to separate these kinematical regions. 
As Fig.~\ref{rhoexp} demonstrates, at $-t\leq 0.6$~GeV$^2$ the cross section is 
sensitive to the screening effects,  while at $-t\geq 0.6$~GeV$^2$ it is 
sensitive to the double scattering contribution. Afterwards one has to study the 
$Q^2$-dependence of  the cross sections  in these kinematic regions.  
 
To identify unambiguously the observed $Q^2$-dependence with the onset of  
CT one should however  impose additional kinematic constraints  based on the fact that  
in lepto--production processes, the longitudinal interaction length  plays an  
important role and has a characteristic $Q^2$-dependence(see e.g. \cite{Gribov}): 
\begin{equation} 
l_c~=~{{2\nu}\over{Q^2~+~m_V^2~-t_{min}}}. 
\label{llenght} 
\end{equation} 
 
An important aspect of the measurements is the ability to separate the effects of  
a changing  longitudinal interaction length from those of color coherence with an 
increase of $Q^2$. This  can be achieved by keeping $l_c$  fixed in a $Q^2$ scan of 
the coherent cross section at a wide range of momentum transfers $t$. 
 
Based on the above discussions one can identify a CT observable as the ratio of  
two differential cross sections at fixed $l_c$ but at different $t$: 
one in the double scattering ($t_1$), and another in the screening ($t_2$) regions,  
\begin{equation} 
R = {d\sigma(Q^2,l_c,t_1)/dt\over d\sigma(Q^2,l_c,t_2)/dt} 
\label{Rt} 
\end{equation} 

\begin{figure}[ht] 
\begin{center} 
\epsfig{width=4.0in,file=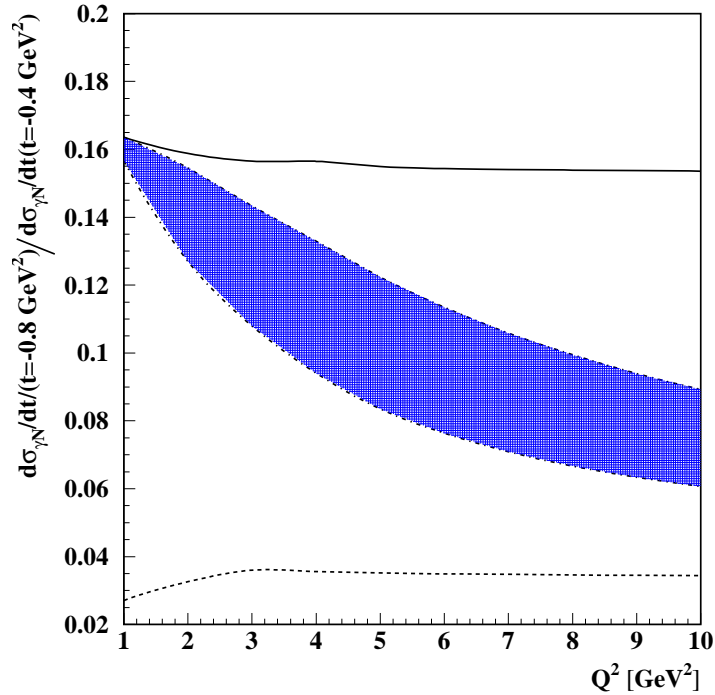}   
\end{center} 
\caption{The ratio $R$ of the cross sections at transferred momenta 
$-t~=~0.4$~GeV/c$^2$, and $-t~=~0.8$~GeV/c$^2$ as a function of $Q^2$.} 
\label{tratio} 
\end{figure} 
 
Figure~\ref{tratio} presents model calculation of the $Q^2$-dependence of 
ratio $R$ for $\rho$ production for $-t_1~=~0.8$ (GeV/c)$^2$ 
and $-t_2~=~0.4$ (GeV/c)$^2$. 
The upper curve is calculated without CT effect, within VMD with a finite  
longitudinal interaction length taken into account~\cite{FKMPSS97}. The lower 
band corresponds to calculations within the quantum diffusion model of CT~\cite{FLFS88} 
with different assumptions for CT~\cite{FPSS98} with respect to the expansion of the PLC and  
its interaction with the spectator nucleon. The upper and the lower limits in  
the band correspond to $\Delta M^2=$ $1.1$ and $0.7$~GeV$^2$ respectively  
(see discussion in Sec.(3.3)).   
 
It is worth noting that for a complete understanding of the coherent production  
mechanism and the formation of the final mesonic states, these measurements should also 
be carried out at $Q^2<1$~GeV$^2$ to allow us to match the theoretical calculations 
with VMD.

\subsubsection {Experimental Objectives}  
 
\begin{figure}[ht] 
\begin{center} 
\epsfig{width=4.0in,file=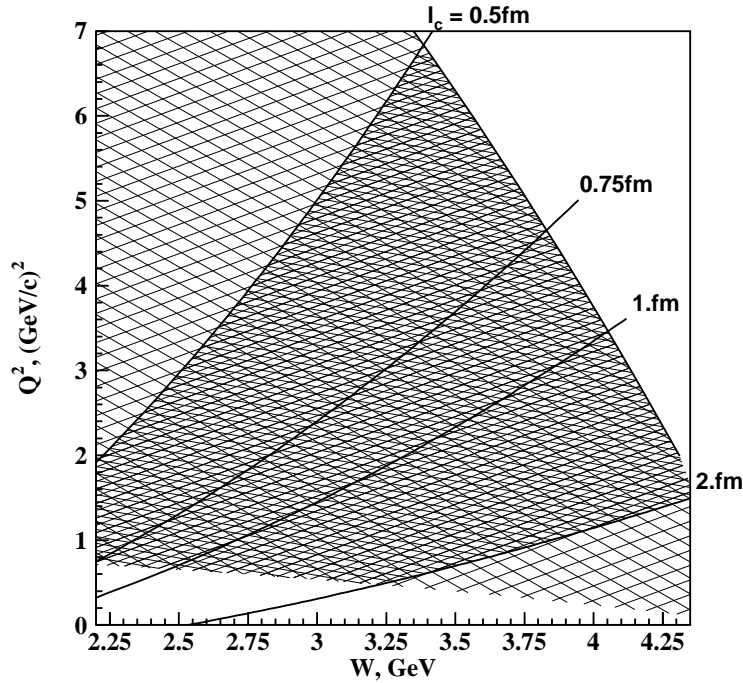}   
\end{center} 
\caption{ 
The accessible range of $Q^2$ and W at $11$~GeV beam energy 
with new design of CLAS. Light shaded region is defined by 
the detection of the scattered electrons in the forward region 
of the CLAS. Dark shaded region is the preferred kinematical 
region for the proposed experiment. Lines represent $Q^2-W$ 
dependence at a fixed longitudinal interaction length.} 
\label{wq2} 
\end{figure}

The experiment studying the reaction of Eqn.~(\ref{vmpro}) will be the part of a broad  
effort to establish the existence of color transparency in QCD at intermediate  
energies. 
A large acceptance detector such as CLAS at Jefferson Lab is an ideal tool for  
conducting such experiments. With a single setting it can simultaneously measure  
the coherent production of all vector mesons in a broad kinematic range. 
Figure~\ref{wq2} shows the accessible kinematical range for an   
$11$~GeV electron beam energy with CLAS++. The lines show the $Q^2-W$ dependence at  
fixed coherent length $l_c$.  The plot  shows that at this energy,  the  
shape of the $t$-dependence can be studied up to $Q^2=5$~(GeV/c)$^2$ at $l_c~\sim~$0.8. 
 
In the final state, the scattered electron and the recoil deuteron will 
be detected together with the decay products of the produced vector meson 
In the case of $\rho^0\to\pi^+\pi^-$, only one of the decay  
pions will be detected, and missing mass technique will be used  
to identify the second one. Additional suppression of the three pion final state 
($\pi^+\pi^-\pi^0$) can be achieved by using a veto on a neutral hit in the CLAS 
calorimeters. For identification of the $\omega$, a neutral 
hit in the calorimeters will be used to suppress the $\rho^0$ 
background. $\phi$ mesons will be identified via their $K^+K^-$ decay, 
detecting one of the kaons. Count rates are estimated with acceptance calculations using 
CLAS++, assuming a luminosity  ${\cal L}=10^{35}\times A/Z$~cm$^{-2}$ 
sec$^{-1}$.  

\begin{figure}[ht] 
\begin{center} 
\epsfig{width=4.0in,file=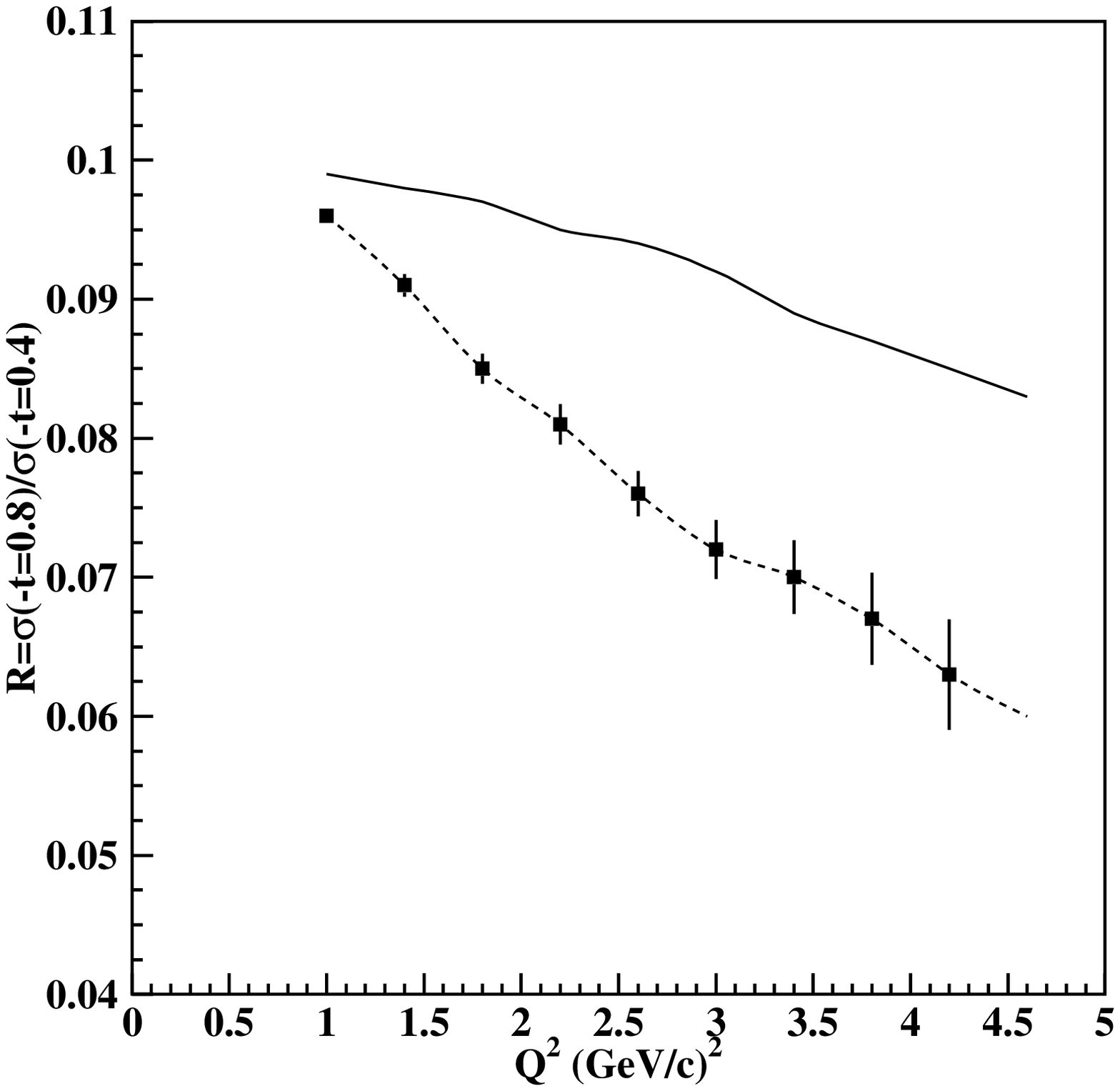}   
\end{center} 
\caption{Expected errors on the ratio of cross sections at transferred 
momenta $0.4$~GeV/c$^2$ and $0.8$~GeV/c$^2$ for 2000 hours of running on CLAS 
with $11$~GeV beam. The kinematics are fixed at $0.75\leq~l_c\leq~0.8$. 
The solid  curve is the calculation of the ratio assuming no color transparency 
effects, the points are with color coherent effect. Events in each point are 
integrated in the bins of $\Delta Q^2~=~0.4$~GeV/c$^2$ and $\Delta l_c~=~0.2$. 
We assume a CLAS++ luminosity of ${\cal L} = 10^{35}\times A/Z$~cm$^{-2}$ sec$^{-1}$.} 
\label{Fig.rates} 
\end{figure}

In Fig.~\ref{Fig.rates}, the expected errors on the ratio of cross sections  
of Eq.~(\ref{Rt}) is presented for the same kinematical conditions  
as in Fig.~\ref{tratio}, with $l_c$ fixed at $0.8$~fm. 
The cross sections are calculated according to Ref.~\cite{FPSS98}. 
The statistical errors correspond to 30 days of beam time. 
This  figure shows that the accuracy of the  
experiment will allow one to  unambiguously verify  
the onset of CT in this region of $Q^2$.

\section{Summary and Discussion} 
 
Quantum Chromodynamics provokes a number of interesting questions related to nuclear 
physics. This review has addresses two of these: 
\begin{itemize} 
\item  What is the quark  nature of nuclei at low temperature and high density?  
 
\item What is the influence of color on hadron-nucleon interactions in nuclei? 
\end{itemize} 
Our central theme is that the use of Jefferson Laboratory, with electron energies  
up to $11$~GeV, will lead to substantial progress in answering these questions.  
 
New studies of deep inelastic scattering by nuclear targets will 
focus on the first question. We discuss how the search for scaling in  
deep inelastic scattering at values of Bjorken $x>1$ will focus on a microscopic  
study of the nature of the quantum fluctuations which briefly transform   
ordinary nuclear matter into a high density system.  Furthermore,  
the measurement of backward going nucleons in coincidence 
with the outgoing electron (denoted as tagging the structure function) will 
lead to disentangling the various models which have been proposed as explaining the 
nuclear EMC effect  and thereby establish a clear signature of quark degrees of freedom  
in the nuclear structure

New studies of the knockout of one or two nucleons by electrons at high momentum 
transfer offer the promise of revealing how   color influences the interaction 
between an ejected color singlet particle and the spectator  nucleons. 
The absence of significant final state interactions, known as color transparency, 
would allow the discovery of a novel new phenomenon in baryon interactions.  
New measurements of the electroproduction of vector mesons in coherent interactions 
with a deuteron target will show how color influences meson-nucleon interactions. 
 
The questions we discuss have been perplexing physicists for more than twenty years. 
The use of Jefferson Laboratory, with its well known high resolution, high duty factor, 
and high luminosity, at an energy of $\simeq 12$ GeV, will  finally  provide the long 
desired answers. 
 
\ack 
This work is supported  by  DOE grants under contracts DE-FG02-01ER-41172, 
DE-FG02-96ER-40960, DE-FG02-96ER40950, DE-FG03-97ER-41014 and W-31-109-ENG-38. 
This work was supported also by the Israel Science Foundation funded 
by the Israel Academy of Science and Humanities. 
We gratefully acknowledge also the support from Jefferson Lab. 
The Thomas Jefferson National Accelerator Facility (Jefferson Lab) is operated  
by the Southeastern Universities Research Association (SURA) under DOE contract  
DE-AC05-84ER-40150.


\section*{References} 
\begin{thebibliography}{02} 
 
\bibitem{Drake} Drake J J \etal 2002 Astrophys. J. {\bf 572} 996 
 
\bibitem{R1} Rock S \etal 1982 \PR C {\bf 26} 1592 
 
\bibitem{R2} Rock S \etal 1992 \PR D {\bf 46} 24 
 
\bibitem{D1} Day D \etal 1979 \PRL {\bf 43} 1142 
 
\bibitem{D2} Day D \etal 1987 \PRL {\bf 59} 427 
  
\bibitem{fsds93} Frankfurt L L, Strikman M I, Day D B, Sargsian M, 1993 \PR D {\bf 48} 2451 
 
\bibitem{liuti93} Liuti S, 1993 \PR C {\bf 47} R1854 
 
\bibitem{Kim01} Egiyan K Sh 2002 Study of Nucleon Short Range Correlations in $A(e,e')X$  
                reactions at $x_B>1$  
                {\it Proc. of the 9th International Conference on the Structure of Baryons} 
 
\bibitem{Pa97} Pandharipande V R, Sick I and deWitt Huberts P K A 1997 \RMP {\bf 69} 981 
 
\bibitem{FS81} Frankfurt L L and Strikman M I 1981 Phys. Rept. {\bf 76} 215 
 
\bibitem{Arrin99} Arrington J \etal 1999 \PRL {\bf 82} 2056  
 
\bibitem{e02019} Arrington J, Day D B, Filippone W B and Lung A (spokespersons) 2002  
                 Inclusive Scattering from Nuclei at $x>1$ and High $Q^2$ with a 6 GeV Beam  
                 {\it Jefferson Lab Proposal E02-019} 
 
\bibitem{Larry01} Weinstein L B 2001 NN Correlations Measured in $^3$He(e,e'pp)n 
                 {\it Proc. of Fifth Workshop on Electromagnetically Induced  
               Two-Hadron Emission}   
 
\bibitem{wp} Cardman L \etal (ed) 2001 The Science Driving the 12 GeV Upgrade of CEBAF  
         {\it Jefferson Lab Report}  
 
\bibitem{EMC1}  Aubert J J \etal (European Muon Collaboration) 1983 \PL B {\bf 123} 275   
 
\bibitem{EMCB1} Bodek A \etal 1983 \PRL {\bf 50} 1431 
 
\bibitem{EMCB2} Bodek A \etal 1983 \PRL {\bf 51} 534  
 
\bibitem{dyth} Bickerstaff R P, Birse M C and Miller G A 1984 \PRL {\bf 53} 2532  
 
\bibitem{dyth1} Ericson M and Thomas A W 1984 \PL B {\bf 148} 191   
 
\bibitem{dyth2} Berger E L 1986 \NP B {\bf 267} 231   
 
\bibitem{DY} Alde D M \etal 1990 \PRL {\bf 64} 2479     
 
\bibitem{FS88} Frankfurt L L and Strikman M I 1988 Phys. Rept.  {\bf 160} 235  
 
\bibitem{FLS} Frankfurt L L, Strikman M I and Liuti S 1990 \PRL  {\bf 65} 1725  
   
\bibitem{Pirner} Gousset T and Pirner H J 1996 \PL B {\bf 375} 349  
 
\bibitem{scqcd} Miller G A 1989 \PR C {\bf 39} 1563  
 
\bibitem{Brown} Brown G E  and Rho M 1991 \PRL  {\bf 66} 2720.

\bibitem{Wilc} Alford M, Rajagopal K and Wilczek F 1998 \PL B {\bf 422} 247   
 
\bibitem{Carter} Carter G W and Diakonov D 2000 \NP B {\bf 582} 571  
 
\bibitem{Rapp} Rapp R, Schafer T, Shuryak E V and Velkovsky M 1998 \PRL  {\bf 81} 53  
  
\bibitem{Dieterich} Dieterich S \etal 2002 to be submitted to \PRL  
 
\bibitem{FSS97} Frankfurt L L, Sargsian M M and Strikman M I 1997 \PR C {\bf 56} 1124  
 
\bibitem{SM} Sargsian M M 2001 Int. J. Mod. Phys. E {\bf 10} 405  
 
\bibitem{Feynman} Feynman R 1972 {\it Photon-Hadron Interactions} (Reading: Benjamin) 
 
\bibitem{FKAS} Frankfurt L L, Kondratyuk A, and Strikman M I 

\bibitem{gdpn} Frankfurt L L, Miller G A, Sargsian M M  and Strikman M I 2000  
               \PRL  {\bf 84} 3045  
 
\bibitem{ISR} Goggi G \etal 1977 \PL B {\bf 72} 265  
 
\bibitem{Stoler} Stoler P 1993 Phys. Rept. {\bf 226} 103  
 
\bibitem{John01}Arrington J 2002 Nucleon Momentum Distributions From a Modified Scaling 
Analysis of Inclusive $e-A$ Scattering {\it Proc. of the 9th International Conference on  
              the Structure of Baryons} 
 
\bibitem{FS78} Frankfurt L L and Strikman M I 1978 \PL B {\bf 76}, 285 
  
\bibitem{FS83} Frankfurt L L and Strikman M I 1983 \NP A {\bf 405} 557 
 
\bibitem{UJ} Ulmer P and Jones M 1994 In-Plane Separations and High Momentum  Structure  
             in d(e,e'p)n {\it Jefferson Lab Proposal E94-004} 
 
\bibitem{KG} Kuhn S E and Griffioen K A (spokespersons) 1994  
            Electron Scattering from a High Momentum Nucleon in Deuterium  
            {\it Jefferson Lab Proposal E94-102} 
 
\bibitem{EGS} Egiyan K Sh, Griffioen K A and Strikman M I (spokespersons) 1994  
            Measuring Nuclear Transparency in Double Rescattering Processes  
            {\it Jefferson Lab Proposal E94-019} 
 
\bibitem{WJKUV}Boeglin W, Jones M, Klein A, Mitchell, Ulmer P and Voutier E 
               (spokespersons) 2001  Short-Distance Structure of the Deuteron and 
                Reaction Dynamics in $^2$H(e,e'p)n   
                {\it Jefferson Lab Proposal E01-020} 
 
\bibitem{Dutta} Dutta D \etal 2000 \PR C {\bf 61} 061602 
 
\bibitem{bert} Liyanage N \etal [Jefferson Lab Hall A Collaboration] \PRL 2001 {\bf 86} 5670 
 
 
\bibitem{Lap} Lapikas L \etal 2000 \PR C {\bf 61} 064325 
 
\bibitem{Zhalov} Frankfurt L, Strikman M and Zhalov M 2001 \PL B {\bf 503} 73  
 
 
\bibitem{quench} Abbott D \etal 1998 \PRL {\bf 80} 5072 
 
\bibitem{BB}BigBite Spectrometer http://hallaweb.jlab.org/equipment/BigBite/index.html 
 
\bibitem{add5:CEBAFpro} Piasetzky E, Bertozzi W, Watson J and Wood S (spokespersons) 2001 
         Studying the Internal Small-Distance Structure of Nuclei Via the Triple Coincidence  
         (e,epN) Measurement {\it Jefferson Lab Proposal E01-015} 
 
\bibitem{FGMSS95} Frankfurt L L \etal \ZP A {\bf 352} 97   
 
\bibitem{scaling} Friedman J I and Kendall H W 1972 Ann. Rev. Nucl. Part. Sci. {\bf 22} 203  
 
 
\bibitem{Bjorken} Bjorken J D and Paschos E A 1969 \PR {\bf 185} 1975  
 
\bibitem{BCDMS} Benvenuti A C \etal (BCDMS Collaboration) 1994 \ZP C {\bf 63} 29    
 
\bibitem{CCFR} Vakili M \etal (CCFR Collaboration) 2000 \PR D {\bf 61} 052003    
 
\bibitem{Arrington} Arrington J \etal 2001 \PR  C {\bf 64} 014602  
 
\bibitem{FDSS} Frankfurt L L, Day D B, Sargsian M M and Strikman M I 1993 \PR C {\bf 48} 2451   
 
\bibitem{FSS90} Frankfurt L L, Sargsian M M and Strikman M I 1990 \ZP A {\bf 335} 431   
 
\bibitem{FS85} Frankfurt L L and Strikman M I 1985 \NP B {\bf 250} 1585   
 
\bibitem{FS80} Frankfurt L L and Strikman M I 1980 \PL B {\bf 94} 216    
 
\bibitem{EPC} Lightbody J W and O'Connel J S 1988 Computers in Physics {\bf May/June} 57  
 
 
\bibitem{EMC2} Benvenuti A C \etal (BCDMS Collaboration) \PL B {\bf 189} 483    
 
\bibitem{EMC3} Ashman J \etal (EMC Collaboration) 1988 \PL  B {\bf 202} 603 
 
\bibitem{EMC3b} Ashman J \etal (EMC Collaboration) 1993 \ZP C {\bf 57} 211    
 
\bibitem{EMC4} Dasu S \etal 1994 \PR D {\bf 49} 5641   
 
\bibitem{EMC5} Gomez J \etal 1994 \PR D {\bf 49} 4348    
 
 
\bibitem{e00101} Arrington J (spokesperson) 2000 A Precise Measurement of the 
                Nuclear Dependence of Structure Functions in Light Nuclei 
                {\it Jefferson Lab Proposal E00-101} 
 
\bibitem{CL} Ciofi~degli~Atti C and Liuti S 1989 \PL B {\bf 225} 215   
 

\bibitem{CLS91}Carlson C E, Lassila K E, and Sukhatme U P 1991 \PL B {\bf 263} 377   
 
\bibitem{CL95} Carlson C E and Lassila K E 1995 \PR C {\bf 51} 364  
 
\bibitem{MSS97} Melnitchouk W, Sargsian M and Strikman M I 1997 \ZP A {\bf 359} 359    
 
\bibitem{Ber78} Berge J P \etal 1978 \PR D {\bf 18} 1367    
 
\bibitem{Efr80} Efremenko V I \etal 1980 \PR D {\bf 22} 2581 
 
\bibitem{AKV} Akulinichev S V, Kulagin S A and Vagradov G M 1985 \PL B {\bf 158} 485 
 
\bibitem{AKVa} Kulagin S A 1989 \NP A {\bf 500} 653    
 
\bibitem{DT} Dunne G V and Thomas A W 1985 \NP A {\bf 446} 437c    
 
\bibitem{ET} Ericson M and Thomas A W 1983 \PL B {\bf 128} 112   
 
\bibitem{ANL} Friman B L, Pandharipande V R, and Wiringa R B 1983 \PRL {\bf 51} 763 
 
\bibitem{ANLa} Berger E L, Coester F and Wiringa R B 1984  \PR D {\bf 29} 398   
 
\bibitem{HM} Jung H and Miller G A 1998 \PL B {\bf 200} 351    
 
\bibitem{MEC} Kaptari L P \etal 1990 \NP A {\bf 512} 684   
 
\bibitem{MECa} Melnitchouk W and Thomas A W 1993 \PR D {\bf 47} 3783    
 
\bibitem{BT} Bickerstaff R P and Thomas A W 1989 \jpg {\bf 15} 1523  
 
\bibitem{MillerSmith} Miller G A and Smith J R 2002 \PR C {\bf 65} 015211 
 
\bibitem{MillerSmith2} Miller G A and Smith J R 2002 \PR C {\bf 65} 055206 
 
\bibitem{HvH} Hugenholtz N M and van Hove L 1958 Physica {\bf 24} 363  
 
\bibitem{KOLTUN} Koltun D S 1998 \PR C {\bf  57} 1210 
 
\bibitem{MST} Melnitchouk W, Schreiber A W and Thomas A W 1994 \PR D {\bf 49} 1183   
 
\bibitem{KPW} Kulagin S A, Piller G and Weise W 1994 \PR C {\bf 50} 1154    
 
\bibitem{GL} Gross, F and Liuti, S 1992 \PR C {\bf 45} 1374 
 
\bibitem{qmc} Guichon P A 1988 \PL B {\bf 200} 235  
 
\bibitem{qmc1} Guichon P A M, Saito K, Rodionov E and Thomas A W 1996 \NP A {\bf 601} 349 
 
\bibitem{qmc2} Blunden P G and Miller G A 1996 \PR C {\bf 54} 359 
 
\bibitem{CLRR} Close F E, Roberts R G and Ross G G 1983 \PL B {\bf 129} 346   
 
\bibitem{JRR} Jaffe R L, Close F E, Roberts R G and Ross G G 1984 \PL B {\bf 134} 449   
 
\bibitem{CLOSE} Close F E \etal 1985 \PR D {\bf 31} 1004    
 
\bibitem{NP} Nachtmann O and Pirner H J 1984 \ZP C {\bf 21} 277    
 
\bibitem{GP} G{\"u}ttner G and Pirner H J 1986 \NP A {\bf 457} 555    
 
\bibitem{JM96} Frank M R, Jennings B K and Miller G A 1996 \PR C {\bf 54} 920    
 
\bibitem{sixquark1} Frankfurt L L and Strikman M I 1975  Proceedings of 10th Winter School 
                    of Physics on Nuclear Physics and Elementary Particle  {\bf 2} 449

\bibitem{sixquark2}Brodsky S J and Chertok B T  1976 \PR D {\bf 14} 3003

\bibitem{sixquark3}Matveev V A and Sorba P 1978 Nuovo Cim.  A  {\bf 45}  257

\bibitem{PV} Pirner H J and Vary J P 1981 \PRL {\bf 46} 1376

\bibitem{CH} Carlson C E and Havens T J 1983 \PRL {bg 51} 261

\bibitem{gm6q} Henley E M, Kisslinger L S and Miller G A 1983 \PR C {\bf 28} 1277  
 
\bibitem{gm6q1} Miller G A 1984 \PRL {\bf 53} 2008     
 
\bibitem{gm6q2} Miller G A 1986 \PL B {\bf 174} 229  
 
\bibitem{gm6q3} Koch V and Miller G A 1985 \PR C {\bf 31} 
 
\bibitem{MULDERS1} Mulders P J and Thomas A W  1984 \PRL  {\bf 52} 1199 
 
\bibitem{MULDERS2}Mulders P J and Thomas A W 1983 J. Phys. G {\bf 9} 1159 
 
\bibitem{GM85}Miller G A 1985 \NP A {\bf 446} 445.

\bibitem{KS85}Kondratyuk L A Shmatikov M 1985 Sov. J. Nucl. Phys. {\bf 41} 141

\bibitem{LS} Lassila K E and Sukhatne U P 1988 \PL B {\bf 209} 343

\bibitem{CL59}Chew G F and Low F E 1959 \PR {\bf 113} 1640 
 
\bibitem{FSSprep} Frankfurt L, Sargsian M and Strikman, {\em in preparation}  
 
\bibitem{FMS93} Frankfurt L, Miller G A and Strikman M 1993 \PL B {\bf 304} 1   
 
\bibitem{Brod94}Brodsky S J, Frankfurt L, Gunion J F, Mueller A H and Strikman M 1994  
                \PR D {\bf 50} 3134   

\bibitem{CFS97} Collins J C, Frankfurt L and Strikman M 1997 \PR D {\bf 56} 2982   

\bibitem{AC99} Abramowicz H and Caldwell A 1999 \RMP {\bf 71} 1275   
 
\bibitem{Ashery} Aitala E M \etal (E791 Collaboration) 2001 \PRL {\bf 86} 4768 
 
\bibitem{Ashery2} Aitala E M \etal (E791 Collaboration) 2001 \PRL {\bf 86} 4773 
 
\bibitem{E665} Adams M R \etal (E665 Collaboration) 1995 \PRL {\bf 74} 1525    
 
 
\bibitem {Brodsky82} Brodsky S J 1982 {\it Proc. of the Thirteenth Int'l Symposium on    
                     Multiparticle Dynamics} (Singapore: World Scientific) p 963    

\bibitem {Mueller82} Mueller A H 1982 {\it Proc. of the Seventeenth Rencontres de Moriond}    
                    vol~1 (Gif-sur-Yvette: Editions Fronti{\`e}res) p~13   
 
\bibitem{FLFS88} Farrar G R, Liu H, Frankfurt L L and Strikman M I 1988 \PRL {\bf 61} 686   
 
\bibitem{FSZ90} Frankfurt L L, Strikman M and Zhalov M 1990 \NP A {\bf 515} 599 
 
\bibitem{jm90} Jennings B K and Miller G A 1990 \PL B {\bf 236} 209   
 
\bibitem{jm90a} Jennings B K and Miller G A 1991 \PR D {\bf 44} 692 
 
\bibitem{jm90b} Jennings B K and Miller G A 1992 \PRL {\bf 69} 3619 
 
\bibitem{jm90c} Jennings B K and Miller G A 1992 \PL B {\bf 274} 442             
 
\bibitem{boffi} Bianconi A, Boffi S and Kharzeev D E 1993 \PL B {\bf 305} 1  
 
\bibitem{BM88} Brodsky S J and Mueller A H 1988 \PL B {\bf 206} 685 
 
\bibitem{DKMT} Dokshitzer Y L, Khoze V A, Mueller A H and Troyan S I 1991 {\it Basics of    
               Perturbative QCD} (Gif-sur-Yvette: Editions Fronti{\`e}res)    

\bibitem{GFMS92} Frankfurt L L, Greenberg W R, Miller G A and Strikman M 1992 \PR C {\bf 46} 2547  
 
 
\bibitem{JRS} Szczepaniak A, Radyushkin A and Ji C 1998 \PR D {\bf 57} 2813  
               
\bibitem{JRSa} Melic B, Nizic B and Passek K 1999 \PR D {\bf 60} 074004 
 
\bibitem{fmsplc} Frankfurt L, Miller G A and Strikman M 1993 \NP A {\bf 555} 752  
 
\bibitem {rp88} Ralston J P and Pire B 1988 \PRL {\bf 61} 1823  
 
\bibitem {bdet} Brodsky S J and de Teramond G F 1988 \PRL {\bf 60} 1924  
 
\bibitem{pire} Jain P, Pire B and Ralston J P 1996 Phys. Rept. {\bf 271} 67   
 
\bibitem{Car88} Carroll A S \etal 1988 \PRL {\bf 61} 1698    
 
\bibitem{FSZ94} Frankfurt L L, Strikman M and Zhalov M 1994 \PR {\bf C50} 2189 
  
\bibitem{BNL98} Mardor Y \etal 1998 \PRL {\bf 81} 5085   
 
\bibitem{BNL01} Leksanov A \etal 2001 \PRL {\bf 87} 212301  
 
\bibitem{jm94} Jennings B K and Miller G A 1993 \PL B {\bf 318} 7  
 
\bibitem{NE18} O'Neill T G \etal 1995 \PL B {\bf B351} 87 
 
\bibitem{NE18a} Makins N C R \etal \PRL {\bf 72} 1986  
 
\bibitem{garrow} Garrow K \etal 2001  {\it Preprint} hep-ex/0109027, \PR C {\em in press} 
 
\bibitem{FMSS} Frankfurt L L, Moniz E J, Sargsian M M, Strikman M I 1995 \PR C {\bf 51} 3435   
 
\bibitem{NSSW} Nikolaev N N, Szczurek A, Speth J, Wambach J,  
               Zakharov B G and Zoller V R 1994 \PR C {\bf 50} 1296  

\bibitem{FS91} Frankfurt L L and Strikman M 1991 Prog. in Part. and Nucl. Phys. {\bf 27} 135   
 
\bibitem{double} Egiyan K S \etal \NP A {\bf 580} 365     
 
\bibitem{fms}Frankfurt L L, Miller G A and Strikman M 1994 Ann. Rev. of Nucl. and  
             Part. Sci. {\bf 44} 501  
 
\bibitem{chiral} Frankfurt L L, Lee T S H, Miller G A and Strikman M 1997 \PR C {\bf 55} 909  
  
\bibitem{GV}Vanderhaeghen M,  Guichon P A  and Guidal M 1999 \PR D {\bf 60} 094017

\bibitem{FKS96} Frankfurt L, Koepf W,  and Strikman M  1996 \PR D {\bf 54} 3194

\bibitem{Lech} Mankiewicz L, Piller G and Weigl T 1998  \EJP {\bf 5} 119

\bibitem{KNNZ}Kopeliovich B Z, Nemchick J, Nikolaev N N and Zakharov B G 1993 
              \PL B {\bf 309} 179 
 
\bibitem{e01107} Ent R, Garrow K (spokespersons) 2001 Measurement of Pion Transparency 
                 in Nuclei {\it Jefferson Lab Proposal E01-107} 

\bibitem{e02110} Hafidi K, Holtrop M, Mustapha B (spokespersons) 2002 $Q^2$-dependence of Nuclear  
                Transparency for Incoherent $\rho^0$ Production 
                {\it Jefferson Lab Proposal E02-110} 
\bibitem{SKKpro}Stepanyan S, Kramer L and Klein F~(spokespersons) 2001 
               Coherent Vector Meson Production of Deuteron 
                {\it Jefferson Lab Proposal E02-012} 

\bibitem{slac71} Anderson R L \etal 1971 \PR D {\bf 4} 3245    
 
\bibitem{overman} Overman I D 1971 PhD. Thesis, SLAC-140, UC-34   
 
\bibitem{Gribov} Gribov V N 1970 Sov. Phys.-JETP {\bf 30} 709 
 
\bibitem{FKMPSS97} Frankfurt L \etal 1997 \NP A {\bf 622} 511 
 
\bibitem{FPSS98} Frankfurt L, Piller G, Sargsian M and Strikman M 1998 \EJP A {\bf 2} 301 
  
\end{thebibliography}
\end{document}